\input phyzzx
\hoffset=0.375in
\overfullrule=0pt
\font\bigfont=cmr17
\centerline{\bigfont Les Microlentille Gravitationnelles}
\bigskip
\centerline{{\bf Andrew Gould}\footnote{1}{Alfred P.\ Sloan Foundation Fellow}}
\smallskip
\centerline{Dept of Astronomy, Ohio State University, Columbus, OH 43210}
\smallskip
\centerline{and}
\smallskip
\centerline{Coll\`ege de France, Paris}
\smallskip
\centerline{e-mail gould@payne.mps.ohio-state.edu}
\bigskip
\centerline{\bf Abstrait}
\def\event{{\'ev\'enement}}
\def\a{{\`a}\ }
\def\psf{{\rm psf}}
\def\eff{{\rm eff}}

\singlespace

	Je pr\'esente une revue des microlentilles gravitationnelles en quatre 
le\c cons.  Dans un premier temps, je discute la th\'eorie des microlentilles 
dans le contexte g\'en\'eral des lentilles, en incluant les effets de la taille
finie des sources, de la parallaxe, et du retard temporel entre les images.  
Dans la deuxi\`eme le\c con, j'analyse les exp\'eriences courantes.  Je montre 
qu'il n'est pas possible d'expliquer les \event s observ\'es avec les \'etoiles
connues, que ce soit en direction du bulbe ou du Grand Nuage de Magellan.  Je 
d\'ecris quelques nouvelles exp\'eriences qui pourraient nous r\'ev\'eler 
les caract\'eristiques des objets d\'etect\'es.  Le troisi\`eme sujet porte sur
la m\'ethode des pixels, invent\'ee pour observer des \event s de 
microlentilles sur des sources non r\'esolues.  Je pr\'esente une analyse 
g\'en\'erale de la m\'ethode, en incluant des probl\`emes d'alignement 
g\'eom\'etrique, photom\'etrique, et de PSF, du rapport signal sur bruit, et de
la nature des informations tir\'ees des observations.  Je pr\'esente un rapport
sur le progr\`es des exp\'eriences actuelles et je discute de deux id\'ees pour
rechercher des objets sombres dans l'amas de Virgo et des ``microlentilles 
extr\^emes'' dans le bulbe.  Pour finir, je parle de l'avenir des 
microlentilles.  J'explique comment on peut les utiliser pour rechercher des 
plan\`etes, pour mesurer les vitesses transverses de galaxies et les vitesses 
de rotation d'\'etoiles g\'eantes, et pour rechercher les objets de la masse 
d'une com\`ete se trouvant de l'autre c\^ot\'e de l'univers.  Ce texte est
publi\'e sous forme de rapport interne du Laboratoire de Physique 
Corpusculaire du Coll\`ege de France (LPC 96-23).  Il inclue une copie des
transparents de ces quatres le\c cons, copie disponible aupr\`es de Dani\`ele
Levaillant: levaillant@cdf.in2p3.fr

\normalspace
\chapter{La th\'eorie de lentille gravitationnelle}

	1) Je vais vous pr\'esenter quatre le\c cons sur les microlentilles 
gravitationnelles.  Un \event\ de microlentille gravitationnelle se produit 
lorsqu'une masse (la lentille) passe pr\`es de la ligne de vis\'ee d'une 
\'etoile d'arri\`ere plan
au loin (la source).  La lentille n'occulte pas la source.  Au lieu de cela,
elle l'amplifie.  Si la lentille est sombre, on peut n\'eanmoins
la d\'ecouvrir gr\^ace \a cet effet.

	Le sujet de la premi\`ere le\c con, aujourd'hui, est la
th\'eorie des microlentilles, et aussi des lentilles en g\'en\'eral.  Ce
sujet est en gros tr\`es simple en dehors de quelques
raffin\'ements. En premi\`ere approximation, on peut d\'ecrire l'effet de 
microlentille classique avec une fonction d'une seule variable.
Les raffinements concernent premi\`erement des petits effets que
peuvent nous donner des informations importantes, et il faut analyser
l'\'ev\'enement attentivement pour comprendre ces effets.  Deuxi\`emement, 
\a peu pr\`es
10\% des \'ev\'enements sont produits par les lentilles binaires et 
ces \event s ne sont pas simples.  Il faut \'etudier la th\'eorie g\'en\'erale
des lentilles pour comprendre les \event s binaires.

	Le sujet de la deuxi\`eme le\c con sera la physique de la Voie 
Lact\'ee et des \'etoiles qui la constituent.  Les exp\'eriences de
microlentille ont trouv\'e plus de 100 \event s.  Ces r\'esultats sont
\'etonnants:  Il n'est pas possible d'expliquer les \event s observ\'es avec
les \'etoiles connues.  Il semble bien qu'il y ait des objets sombres aussi
bien dans le bulbe central de la galaxie que dans le halo.  Mais 
probablement, ces
objets ne sont pas de m\^eme nature.  Dans le bulbe, ils ont des masses
autour d'un dixi\`eme d'une masse solaire.  Ils pourraient \^etre des
naines brunes.  Dans le halo, ils paraissent avoir des masses proche d'une
demi-masse solaire, comme les naines blanches.  Mais, probablement ces objets
ne sont pas de naines blanches.  Je discuterai quelques exp\'eriences qui
pourraient r\'esoudre ces questions.

	La troisi\`eme le\c con portera sur la m\'ethode des pixels.
Pour tous les \event s observ\'es jusqu'\`a maintentent, les sources
sont dans le bulbe ou dans le Grand Nuage de Magellan (LMC).  Il y a beaucoup
d'\'etoiles r\'esolues dans ces champs, peut-\^etre 50 millions dans le bulbe
et 20 millions dans le LMC.  Il y a aussi quelques millions d'\'etoiles 
r\'esolues
dans le Petit Nuage de Magellan (SMC), mais, c'est tout.  On ne conna\^\i t
pas d'autre champs avec un grand nombre des \'etoiles r\'esolues. Il existe
un certain nombre de 
champs avec beaucoup d'\'etoiles non r\'esolues, comme la grande
galaxie dans Androm\`ede, M31.  Les \event s de microlentille vers ces champs
nous donneraient d'informations tr\`es int\'eressantes, si nous pouvions les
observer.  Malheureusement, les m\'ethodes classique de microlentilles sont 
inutilisable pour ces champs.  
Cependant, une nouvelle m\'ethode, la m\'ethode des
pixels cr\'e\'ee par AGAPE ici et par un autre groupe \`a New York, est
utilisable pour ces champs.  De plus, on peut appliquer la m\^eme m\'ethode 
aux champs avec les \'etoiles r\'esolues et observer ainsi certains effets
fins que j'appelle ``microlentilles extr\^emes''.

	La quatri\`eme le\c con parlera de l'avenir.  L'id\'ee de 
microlentille est tr\`es jeune.  Il n'y a que six ans, tous le monde pensaient
que l'effet de microlentille \'etait {\it un} seul effet et qu'il ne pouvait
servir que un seul but: chercher des objets sombres dans le halo.  
Aujourd'hui, 
il y a beaucoup de projets pour appliquer l'effet de microlentille et, 
sans do\^ute,
demain il y en aurait encore plus.  Je discuterai quelques id\'ees, comme
mesurer les rotations des \'etoiles g\'eantes, ou chercher pour des
petits objets des com\`etes au confins de l'univers.

	2)  Laplace et Einstein nous disent ``La gravitation courbe la
lumi\`ere''.  En utilisant la th\'eorie classique de la gravitation, on
peut facilement arriver \`a cette formule, ou $\alpha$ est l'angle de
la d\'eviation, $M$ est la masse (suppos\'ee ponctuelle), $b$ est 
le param\`etre de l'impact, $G$ est la constante de Newton, et
$c$ est la vitesse de la lumi\`ere.  Einstein a rectifi\'e cette formule
en changant le ``2'' en ``4''.  Comment sait-on que Einstein
avait raison?  D\`es 1919, Eddington fait son exp\'erience
fameuse  pendant une \'eclipse de soleil.  Cet exp\'erience a test\'e
la th\'eorie pour une masse solaire et pour $b$ \'egal \`a un rayon de
soleil.  Avec l'effet de microlentille, on veut \'eprouver la th\'eorie pour
une masse solaire et pour $b$ \'egal \`a une unit\'e astronomique 
(UA).  L'exp\'erience d'Hipparcos a test\'e la th\'eorie pour
ces valeurs des param\`etres. Cette formule est valable quand l'observateur est
pr\`es de la masse.  Pour le cas d'Eddington ($\eta$ tr\`es petit), elle
r\'eduit en cette formule l\`a.  Hipparcos a deux t\'elescopes 
s\'epar\'es par 58 d\'egres pour mesurer la parallaxe absolue, pas la parallaxe
relative aux \'etoiles au loin au m\^eme champ.  
Donc, la lumi\`ere re\c cue par un
t\'elescope est courb\'e plus que la lumi\`ere re\c cue par l'autre.  La
grandeur de cet effet est le carr\'e du rapport de la vitesse de la terre \a
la vitesse de la lumi\`ere, ou $10^{-8}$.  C'est 
\'equivalent \`a 2 milli-arcsec.  Puisque l'exp\'erience a mesur\'e
les parallaxes avec une pr\'ecision de 2 mas pour chaque \'etoile,
et puisqu'il y avait cent mille \'etoiles, elle a mesur\'e l'effet
gravitationnel avec une pr\'ecision moins de 1\%.  Les effets
des microlentilles gravitationnelles sont donc tr\`es bien compris.

	3) Dans cette figure, la source est rouge, la lentille est noire,
et l'observateur est bleu.  La lumi\`ere est d\'efl\'echie d'un angle
$\alpha$.  $\theta_S$ est l'angle entre la source et la lentille, vu par
l'observateur.  Il y a
deux images, une image \`a chaque c\^ot\'e de la lentille.  Les angles
entre la lentille et les images sont $\theta_{I+}$ et $\theta_{I-}$.
La distance entre l'observateur et la lentille est $D_l$, 
la distance entre l'observateur et la source est $D_s$, et
la distance entre la lentille et la source est $D_{ls}$.
Dans cette \'equation, l'angle ext\'erieur ($\alpha$) est \'egal au total
des angles internes.  Ces angles sont $q/D_l$ et $q/D_{ls}$.
Mais $q$ est \'egal \`a $D_{l}$ multipli\'e par $\theta_I-\theta_S$.
Je d\'efinis le rayon d'Einstein $\theta_e$ comme \c ca et alors, l'\'equation
devient simple.  Les deux solutions sont plus ou moins $x_\pm$, 
o\`u $x_\pm$ sont comme \c ca.
Maintenant, je commence \a travailler \a l'int\'erieur de l'anneau d'Einstein.  
La position
de la source est simplement $x$, et les positions des images sont $x_\pm$.
Tous les angles sont normalis\'es au rayon d'Einstein.

	4) Imaginons une \'etoile carr\'ee.  Chaque c\^ot\'e est $z$.
Les positions de la source et de l'image sont $x$ et $x_+$.  Alors,
la longueur tangentielle est mulipli\'ee par $x_+/x$.  
Pareillement,
la longueur radiale est multipli\'ee par $\partial x_+/\partial x$.  
Par le th\'eor\`eme de Liouville, la densit\'e de flux par unit\'e d'angle
solide est la m\^eme pour la source et pour l'image.  
L'amplification est donc \'egale au rapport de la
surface angulaire de l'image \a la surface angulaire de la source, 
et elle est donn\'ee par cette formule pour chaque image.  
L'amplification
totale est une peu compliqu\'ee, mais pour petit $x$, elle est simplement
$1/x$.  Aussi, la diff\'erence entre les deux amplifications
est exactement 1.  Cette derni\`ere formule ici est utile pour les probl\`emes
o\`u intervient l'interf\'erence entre les deux images, que je discuterai
dans la quatri\`eme le\c con.

	5) Quand l'observateur, la lentille, et la source sont tous les trois
align\'es, l'image est un anneau, et elle est appel\'ee 
``l'anneau d'Einstein''.  
Le rayon angulaire de l'anneau est appel\'e ``le rayon angulaire d'Einstein'',
$\theta_e$, et le rayon physique est appel\'e simplement
``le rayon d'Einstein'', $r_e$.  Quand les trois points ne sont pas align\'es,
le sym\'etrie est cass\'e, et il y a deux images.  L'amplification est
toujours donn\'ee par le rapport de la surface de l'image \`a la surface de
la source.  Si la lentille n'est pas trop proche de la source, l'amplification
est donn\'ee par cette formule.

	6) Pour des param\`etres typiques, le rayon angulaire d'Einstein va
d'un dixi\`eme milli-arcsec \`a un milli-arcsec.  Le rayon d'Einstein est de
quelques unit\'es astronomiques.  A l'int\'erieur de l'anneau d'Einstein, 
l'amplification est approximativement $1/x$.

	7) On ne sait pas si une source est amplifi\'ee parce que on ne 
conna\^\i t
pas le flux sans amplification.  Donc, il faut voir un \event, pas seulement
une amplification.  Un \event\ est d\'ecrit par 3 param\`etres,
$t_0$, $\beta$, et $t_e$.  $t_e$ est le temps mis par la source
pour parcouvrir un rayon d'Einstein.  $\beta$ est le param\`etre d'impact
divis\'e par le rayon d'Einstein, et $t_0$ est l'instant o\`u l'amplification
est la plus grande.  La s\'eparation entre la lentille et la source est 
donn\'ee par le th\'eor\`eme de Pythagore.  L'amplification est une fonction 
de cette s\'eparation seulement.  Donc, toutes les courbes de lumi\`ere 
ressemblent \a \c ca.

	8) Le principe de l'effet de microlentille est montr\'e ici.
L'observateur choisit une source, et peut alors d\'ecouvrir une lentille
entre lui et la source.  Si la source est dans le LMC, la lentille peut \^etre
dans le halo de la Voie Lact\'ee, mais elle peut aussi \^etre dans le disque
de la Voie Lact\'ee ou dans le LMC lui-m\^eme.  Si la source est dans une autre
galaxie comme M31, la lentille peut aussi \^etre dans cette galaxie.
L'observateur peut choisir un quasar.  En ce cas, la lentille peut \^etre
dans une galaxie tr\`es lointaine entre l'observateur et le quasar ou,
peut-\^etre, la lentille n'est pas dans une galaxie, mais entre des galaxies.

	9) La probabilit\'e qu'une \'etoile soit amplifi\'ee par une lentille
est appel\'ee $\tau$, ou ``la profondeur optique''.  Cette probabilit\'e
para\^\i t tr\`es compliqu\'ee parce qu'elle est le total des sectionnes
efficaces de tous les disques d'Einstein de l'observateur \`a la source.
Dans cette formule, $M_i$ est la masse de l'objet du type $i$, et $n_i(D_l)$
et la densit\'e num\'erique des objet de ce type.  Mais, cette expression 
ici est simplement la densit\'e de masse totale.  
Alors, l'\'equation pour la probabilit\'e est une
int\'egrale simple.  De plus, le th\'eor\`em du viriel nous permet \'evaluer
approximativement cette int\'egrale en termes de la vitesse caract\'eristique
du syst\`eme, comme la vitesse de rotation d'une galaxie.  Par exemple, 
on peut estimer que cette probabilit\'e $\tau$ est de 
un par million pour le bulbe et halo de la Voie Lact\'ee, et de un par
milliard pour un amas globulaire d'\'etoiles.

	10) Il y a un \event\ par million d'\'etoiles observ\'ees.
Dans ce million d'\'etoiles, il y a peut-\^etre mille \'etoiles
variables.  Cependent, puisque la courbe de lumi\`ere des \event s de 
microlentille est d\'ecrite
par seulement trois param\`etres, on peut distinguer les \event s de
microlentille des \'etoiles variables.  Mais pour la m\^eme raison, 
pour chaque \event, on ne peut d\'eterminer que
trois quantit\'es: $t_0$, $\beta$, et $t_e$.  Deux de ces trois param\`etres
sont compl\`etement inutiles.  Ils ne nous inform que sur la g\'eom\'etrie de
l'\event, et pas sur les caract\'eristiques de la lentille.  Le troisi\`eme
param\`etre $t_e$ est utile, mais il est form\'e de quatre autre quantit\'es.
$t_e$ est le rayon d'Einstein divis\'e par la vitesse, $v$.
Le rayon d'Einstein est un m\'elange de la masse et la distance.  La vitesse,
$v$, est une combinaison de la vitesse de la source et de la vitesse de
la lentille.  Donc, la relation entre $t_e$ et les caract\'eristiques
physiques de la lentille est tr\`es indirecte.

	11) Face \a ces difficult\'es, il y a deux m\'ethodes possible.
La premi\`ere: modeliser les distributions de la densit\'e et des vitesses
des sources et des lentilles, et utiliser la distribution de $t_e$ 
observ\'ee pour mesurer la fonction de masse des lentilles. Avec cette 
m\'ethode,
on peut estimer aussi la probabilit\'e $\tau$.  L'autre m\'ethode consiste \a
obtenir plus d'informations pour chaque \event.  Je vais discuter les
deux m\'ethodes bri\`evement maintentant, et je reviendrais plus en d\'etail
sur ce sujet dans la deuxi\`eme le\c con.

	12) Ici, je montre \`a peu pr\`es 50 \event s
observ\'es par MACHO et OGLE vers le bulbe galactique.  Les dur\'ees sont
typiquement 10 jours.  

	13) Cheongho Han et moi-m\^eme
avons modelis\'e les distributions de la densit\'e et de la vitesse
pour pr\'edire la distribution de $t_e$. Puis,
nous avons compar\'e cette distribution avec la distribution de $t_e$ 
observ\'ee.  Si nous utilisons la fonction de masse mesur\'ee avec le
Hubble Space Telescope (HST), les deux distributions sont en d\'esaccord.
L'accord est meilleure si nous utilisons une loi de puissance pour la fonction
de masse.  Mais cette loi elle-m\^eme est en d\'esaccord profond avec 
les observations de Hubble.

	14) L'autre m\'ethode, chercher \a obtenir plus d'informations, 
est tr\`es difficile.  Je vous ai dit qu'on peut mesurer trois quantit\'es
pendant un \event, mais deux de ces trois param\`etres sont inutiles.
Le temps de l'\event\ et le param\`etre d'impact nous disent la
g\'eom\'etrie de l'\event\ et pas les caract\'eristiques de la lentille.  
Alors, nous appliquons la sagesse orientale...

	15)...et lancons un satellite de parallaxe.  Ce satellite peut
convertir les quantit\'es inutiles en des quantit\'es utiles.  Comment?
En moment de la premi\`ere figure, l'anneau d'Einstein recouvre la terre.  
Puis, il
recouvre le satellite.  Les courbes de lumi\`ere vues de la terre et
du satellite ont des $t_0$ et des $\beta$ differents.

	16) On peut reconstruire l'anneau d'Einstein en comparant les
deux courbes de lumi\`ere.  Ici, l'\event\ arrive quatre jours plus t\^ot 
vu du satellite que vu de la terre.  Puisque $t_e$ est 20 jours, la
difference est $0,2$ d'un rayon d'Einstein.  De plus, l'amplification est plus 
petite.  On trouve que les param\`etres d'impact sont 0,4 et 0,8 pour la terre
et le satellite, respectivement.  En conclusion, la diff\'erence des postions
de la source vues de la terre et du satellite est de 0,2 rayon
d'Einstein dans une direction et de 0,4 rayon dans l'autre direction,
soit une diff\'erence totale de $\Delta x$ de 0,45.  Mais, on conna\^\i t
la distance physique
entre le satellite et la terre.  Donc, on peut d\'eduire le longueur du rayon
d'Einstein.  La figure sup\'erieur ressemble \`a une figure que
je vous ai montr\'ee au commencement de cette le\c con.  Mais, maintenant
il y a une autre ligne, le rayon d'Einstein projet\'e sur le
plan de l'observateur.
On peut mesurer ce rayon avec un satellite de parallaxe.  Si on pouvait aussi
mesurer le rayon angulaire d'Einstein, $\theta_e$, la 
g\'eom\'etrie de cette figure serait compl\`etement d\'etermin\'ee.  On 
pourrait donc mesurer la masse, la distance, et la vitesse de la lentille parce
qu'il y a trois quantit\'es inconnues, $M$, $D_l$, et $v$, et il y aurait trois
quantit\'es connues, $\tilde r_e$, $t_e$, et $\theta_e$.  Comment peut-on
mesurer $\theta_e$?

	17) Si on pouvait r\'esoudre les deux images, on mesurerait directement
le rayon angulaire d'Einstein.  C'est possible pour les rayons les plus grands 
par interf\'erometrie.  Cependant, les \event s les plus fr\'equents et aussi
les plus int\'eressants ont des petit rayons 
d'Einstein. Comment peut-on mesurer $\theta_e$
pour ceux-ci?  Si de plus la source est une \'etoile g\'eante, le rayon
angulaire de la source n'est pas beaucoup plus petit que le rayon angulaire
d'Einstein.  Il y a donc une probabilit\'e significative pour que la lentille
passe \`a travers la source.  La courbe de lumi\`ere d\'evierait alors
de la courbe normale.  Habituellement, l'amplification est une fonction de $x$
seulement.  Mais, quand la lentille passe \`a travers la source,
l'amplification change.  Pour une source ponctuelle, les surfaces brunes
des images dans la figure seraient toutes les m\^emes.  Notons que ces
figures ne repr\'esentent pas une \'evolution.  Elles sont des \event s 
diff\'erents avec des rayons angulaire d'Einstein diff\'erents.
Mais, en r\'ealit\'e, quand la source est \'etendue, la surface est plus grande
quand la lentille est proche de la source, et plus petite quand la lentille
est \a l'int\'erieur de la source.
Sur la figure du bas, je montre une 
\'etoile avec un centre bleu et un bord rouge.  Les surfaces bleu et rouge
sont \'egal.  Mais pour les images, les surfaces rouges sont plus grandes.
Une \'etoile comme \c ca para\^\i trait plus rouge pr\`es du pic de l'\event.
Une \'etoile avec un centre bleu para\^\i t bizarre, n'est-ce pas?
Mais toutes les \'etoiles ont des centres bleus parce qu'elles sont plus 
sombres sur leurs bords et ceci plus pour le bleu que pour le rouge.  Il est 
possible d'observer cet effet quand la lentille est plus proche de la source
que deux rayons de la source.

	18) C'est un \event\ de l'exp\'erience MACHO o\`u on peut voir
le premier effet.  Cette courbe a \'et\'e construite en supposant une source
ponctuelle.  Pour l'autre courbe, on suppose que la lentille passe \`a
travers la source.  La deuxi\`eme courbe est meilleure.

	19) Pour finir, je discuterai des lentilles non-ponctuelles.
Pour simplifi\'e, je suppose un sym\'etrie cylindrique.  
La lumi\`ere est courb\'ee seulement par la masse \a l'int\'erieur de
sa trajectoire.  Donc, $\alpha$ est une fonction de $M(\theta_I)$,
pas seulement $M$.  A part cela, l'\'equation est la m\^eme que pour une
masse ponctuelle.  N\'eanmoins, ce changement nous oblige \`a chercher une 
solution graphique au lieu d'une solution analytique.  La courbe bleue, $T_M$,
est le membre de gauche de cette \'equation.  Les lignes vertes, $T_D$, sont le
membre de droite, avec deux $\theta_S$ diff\'erents.  Les deux $\theta_S$
repr\'esentent les extr\'emit\'es de la source.  Les lignes vertes coupent
l'axe horizontal \`a la s\'eparation entre la masse et la source.  Donc,
la fl\`eche bleue repr\'esente la grandeur angulaire de la source.  Les
points o\`u les lignes vertes coupent la courbe bleue sont les solutions de  
cette \'equation.  Il y a donc trois
images et les fl\`eches rouges repr\'esentent les positions et les grandeurs
des images.  Les directions des fl\`eches repr\'esentent les orientations des
images.  On peut calculer l'amplification comme pour une masse
ponctuelle.  Elle est le produit de deux termes: l'un est 
$\theta_I/\theta_S$, et l'autre est la grandeur de la fl\`eche rouge
divis\'e par la grandeur de la fl\`eche bleue.  Quand une ligne verte
approche de la tangente de la courbe bleue, l'amplification d'une source
ponctuelle devient infinie.  De plus, \`a ce point, deux images se fusionnent
et alors disparaissent.

	20) Les positions de la source o\`u l'amplification est infinie
sont ``les caustiques'', et les positions de l'image sont
``la courbe critique''.  Pr\`es d'une caustique, l'amplification peut \^etre
modelis\'ee simplement parce que la ligne verte est lin\'eaire et la courbe 
bleue
est ordinairement, du second degr\'e.  La s\'eparation des deux images
$(\Delta\theta_I)$ varie donc comme la racine carr\'ee de $\Delta \theta_S$, la
s\'eparation entre la position de la source et la caustique.  On voit aussi
que l'amplification varie comme la racine carr\'ee de l'inverse de 
$\Delta\theta_S$ parce que l'amplification se comporte comme l'inverse de
l'angle entre
la ligne verte et la courbe bleue.  Pour finir, les retards entre
les images sont donn\'ees par les int\'egrales montr\'ees comme surfaces
ombrag\'ees.  L'ordre temporel des images est image-1, image-3, image-2.
La surface noire signifie un retard negatif, ou un avance.  Pourquoi ces
int\'egrales sont-elles \'egales aux retards.  C'est parce que
$T_M$ est la d\'eriv\'ee par rapport \a $\theta_I$ du retard gravitationnel, et
$T_D$ est la d\'eriv\'ee du retard g\'eom\'etrique.  Pr\`es d'une caustique,
le retard se comporte comme la s\'eparation des images au cub\'e.

	21)  Pour une masse ponctuelle, l'\event\ a cette allure.  Il n'y a
pas de caustique.  Mais pour un \event\ avec une caustique, la source
se d\'eplace comme \c ca, et tout d'un coup, apparaissent deux images et 
l'amplification devient infinie.  Puis, l'amplfication retombe comme 
la racine carr\'ee du temps.

	21) Maintenant regardons des \event s r\'eels. Tous sont normaux.  
Ils se produisent \a des temps $t_0$ differents.  Les maxima de l'amplication 
sont differents parce que les $\beta$ sont differents.  Les dur\'ees $t_e$ 
sont differentes.  Mais tous les \event s sont essentiellement les m\^emes.

	22) Pour ces \event s aussi...except\'e celui-ci.  Pourquoi est-il
different?

	23) Cet \event\ est une microlentille binaire.  Mon \'etudiant
Scott Gaudi a fait ces figures pour expliquer des \event s binaires.
Les deux lentilles binaires sont ici.  La source est rouge.  Quand la source
est au dehors de la caustique, il y a trois images, montr\'ees en vert.
Ici, la source passe \`a travers la caustique, et deux nouvelles images
apparaissent, un \a l'ext\'erieur de la courbe critique, et 
l'autre \a l'int\'erieur.  Les images sont toutes deux grandes, l'amplification
est donc grande.  Mais, quand la source est au dedans de la caustique, les
images se contractent, et l'amplification tombe.

	24) Ici, il y a un \event\ binaire avec deux caustiques comme 
l'\event\ r\'eel que je vous ai montr\'e.  Mais cet \event\ a aussi
un autre maximum pr\`es d'un point de rebroussement.  Nous verrons un
\event\ r\'eel comme \c ca dans la deuxi\`eme le\c con.

	25) Les conclusions.

\chapter{Les microlentilles et la physique des \'etoiles}

	1) Je vais vous pr\'esenter la deuxi\`eme de quatre
le\c cons sur les microlentilles gravitationnelles.  La premi\`ere le\c con
a couvert la th\'eorie des microlentilles, et je supposerai que vous \^etes
d\'ej\a experts sur ce sujet.  Aujourd'hui, je vais appliquer la th\'eorie aux
exp\'eriences en cours. Le r\'esultat principal est qu'on ne peut pas expliquer
les \event s observ\'es avec les \'etoiles connues.  Il semble bien qu'il y
ait des objets sombres aussi bien dans le bulbe galactique que dans le halo.

	2) Jusqu'\a maintenant, deux groupes recherchent des \event s
de microlentille en observant les \'etoiles du Grand Nuage de Magellan
(aussi appel\'e LMC pour Large Magellanic Cloud en anglais) qui est \a une
distance de 50 kpc et dont la position dans le ciel est de 33 degr\'es 
au sud du plan 
de la Voie Lact\'ee.  Les exp\'eriences sont EROS (Exp\'erience Recherche
des Objets Sombres) et MACHO (Massive Compact Halo Object) qui veut dire
approximativement la m\^eme chose en anglais.  Puisque la ligne de vis\'ee
en direction du 
LMC ne traverse le disque de la Voie Lact\'ee que sur un faible \'epaisseur, 
on n'attend
que peu des \event s dus aux \'etoiles dans cette direction.
C'est donc une bonne direction \a rechercher des objets sombres dans le halo.
On appelle fr\'equemment, 
les objets sombres ``Machos''.
Trois groupes  recherchent des \event s de microlentille en observant les 
\'etoiles du bulbe de la Voie Lact\'ee qui est \a une distance de 8 kpc.
Ce sont MACHO, OGLE (Exp\'erience Optique de Lentille 
Gravitationnelle), et DUO (Objets Invisibles du Disque).
Puisqu'il y a beaucoup d'\'etoiles \a cette direction, on attend
 beaucoup d'\event s dus \a ces \'etoiles.

	3) La question la plus importante est: ``Est-il possible
d'expliquer les \event s observ\'es avec les \'etoiles connues''.
Pour r\'epondre \a cette question, il faut mesurer la profondeur optique
observ\'ee
$\tau_{\rm obs}$, et la comparer \a la profondeur optique dus aux
\'etoiles, $\tau_*$.  Si $\tau_{\rm obs}$ est plus grande que
$\tau_*$, il y a des objets sombres.  Si $\tau_{\rm obs}$ est 
approximativement \'egale \a $\tau_*$, alors il est encore possible
qu'il y a des objets sombres et il faut analyser les \event s plus 
attentivement.  Si $\tau_{\rm obs}$ \'etait plus petite que
$\tau_*$, il y aurait 
un probl\`eme tr\`es grave.  Cette \'equation donne 
$\tau_*$ en termes de $\rho_*(D_l)$, la densit\'e de masse
des \'etoiles en fonction de la distance.  Pour \'evaluer cette \'equation,
il faut conna\^\i tre $\rho_*(D_l)$, que l'on peut \'ecrire $\rho_0$ fois
$F(D_l)$, o\`u $\rho_0$ est la densit\'e\ pr\`es du soleil, et
$F(D_l)$ est la densit\'e relative en fonction de la distance.  
On peut alors compter les \'etoiles au voisinage du soleil et pour chaque type,
multiplier la densit\'e num\'erique par la masse.  Malheureusement, on
mesure les luminosit\'es des \'etoiles, pas leurs masses.  N\'eanmoins, on
peut d\'eterminer une relation entre la couleur et la masse et puis estimer
les masses.

	4)  La densit\'e num\'erique est bien mesur\'e pour les \'etoiles
brillantes, mais il est tr\`es difficile de la mesurer pour les
\'etoiles faibles.  R\'ecemment, nous avons utilis\'e le Hubble Space
Telescope pour d\'eterminer les densit\'es de tous les types des \'etoiles.
Pour chaque \'etoile, je montre la distance au-dessus du plan de la Voie 
Lact\'ee et la magnitude absolue.  La ligne ``b'' repr\'esente la limite
des observations pour les champs typiques, mais les limites sont plut\^ot
``a'' ou
``c'' pour plusieurs des champs.  On voit que la densit\'e tombe pour
les \'etoiles faibles m\^eme dans les zones o\`u l'on n'est pas limit\'e
par l'observation.

	5) Nous avons construit une fonction de luminosit\'e pour d\'ecrire
ces donn\'ees, montr\'ee
avec les triangles verts et l'avons compar\'ee \a celles qu'on trouve dans
la lit\'erature.
On voit que notre solution et celle de Stobie (les cercles oranges)
sont parfaitement d'accord.  Dans 
une estimation ant\'erieure, Wielen avait
trouv\'e  beaucoup plus d'\'etoiles faibles, mais avec une statistique pauvre.
Pour cette raison, certaines personnes ont pens\'e que ces \'etoiles \'etaient
peut-\^etre encore plus nombreuses.  Mais, c'\'etait le cas, 
on pourrait facilement les voir dans les champs du t\'elescope Hubble.

	6) En utisilant une relation empirique
entre la masse et la luminosit\'e, on
peut construire une fonction de masse.  Le pic de cette fonction
est \a une demi-masse solaire (en log de la masse) ou \a une quart de
masse solaire
(en masse).  La fonction para\^\i t remonter au dernier point, mais on
ne sait pas si cette ascension continue pour des masses plus petites, celles
des naines brunes.  On esp\`ere que les microlentilles nous donneront 
des informations sur cette question.

	7) On peut aussi utiliser les \'etoiles observ\'ees par Hubble pour
mesurer leur densit\'e 
en fonction de $z$, de leur distance au-dessus du plan de la Voie
Lact\'ee, et donc d\'eterminer la masse d'une colonne d'\'etoiles
avec une section d'un pc${}^2$: 27 masse solaires.  Les
r\'esultats les plus importants pour nous sont que $\tau_*$ est moins
que $10^{-8}$ dans la direction du LMC et approximativement 
$5\times 10^{-7}$ vers le bulbe.

	8) En plus du disque, il y a une autre population des \'etoiles,
le halo stellaire.  Sa distribution spatialle est approximativement
sph\'erique.  C'est \a dire qu'elle est presque le m\^eme que celle du 
halo sombre. 
On veut donc estimer la fraction de la masse du halo sous forme d'\'etoiles.
Il y a 15 ans, on croyait que le halo stellaire contenaient beaucoup 
d'\'etoiles
faibles.  Il y a 5 ans, Richer \& Fahlman ont pr\'etendu que la fonction
montait vers les luminosit\'es faibles.  
Cependant, plus r\'ecemment, Dahn et al.\ ont
montr\'e qu'il y a peu d'\'etoiles faibles dans le halo stellaire.
Ce r\'esultat implique que $\tau_*$ du halo stellaire est moins que 
$10^{-8}$.

	9) Si Dahn et al.\ 
avaient tort, on verrait beaucoup d'\'etoiles faibles 
dans ``la zone du halo'' du Hubble Deep Field qui a \'et\'e observ\'ee il y 
a 6 mois.  En fait, on n'en a trouv\'e aucune.

	10) Jusqu'\a l'ann\'ee derni\`ere, personne n'avait mesur\'e la
fonction de luminosit\'e de la s\'equence principale du bulbe.  Seules les
\'etoiles g\'eantes et sous-g\'eantes avaient \'et\'e mesur\'ees.  Depuis,
Light et al.\ ont utilis\'e Hubble pour la d\'eterminer jusqu`\a une
magnitude absolue $I$ \'egale 8.  Elle est presque la m\^eme
que la fonction de luminosit\'e du disque.  On peut donc faire
l'hypoth\`ese
que les deux fonctions sont les m\^emes aussi pour les \'etoiles faibles.

	11) Comparons $\tau_*$ et $\tau_{\rm obs}$ pour le bulbe et le
LMC.  Toutes les valeurs sont \a multipli\'ees par $10^8$.  Pour les
observations en direction du bulbe, la profondeur optique dus aux
\'etoiles est 150, dont 50 venant du disque et 100 du bulbe lui-m\^eme.
On voit que $\tau_{\rm obs}$ est un peu plus grande.  Vers le
LMC, la contribution des \'etoiles du disque et du halo stellaire
est inf\'erieure \a deux,
contre 45 si le halo \'etait form\'e enti\`erement
de Machos.  Les \'etoiles du LMC lui-m\^eme contribuent aussi \a la
profondeur optique.  Personne n'a mesur\'e la fonction de masse du LMC
mais plusieurs raisonnements montrent que cette contribution n'est
pas grande.  En fait, $\tau_{\rm obs}$ est beaucoup plus grande
que $\tau_*$.

	12) Pendant la derni\`ere le\c con, je vous ai d\'ej\a montr\'e
la distribution de dur\'ee des \event s observ\'es vers le bulbe par
MACHO et OGLE.  Est-il possible d'expliquer ces \event s avec les
\'etoiles connues?

	13) Zhao, Spergel, \& Rich disent que oui.  Ils reproduisent bien 
la distribution observ\'ee de $t_e$.  Ici, la ligne rouge repr\'esente
les \event s dus aux \'etoiles du bulbe, et la ligne
bleu r\'epresente ceux dus au disque et au bulbe ensemble.  Mais leur
fonction de masse est en d\'esaccord avec celle mesur\'ee.  Ils ont
suppos\'e qu'il n'y a aucune \'etoiles de masse plus grande que
0,6 masse solaire bien que les observations de Light et al.\ montrent
qu'il y en a beaucoup en fait.

	14) Si nous restreignons notre attention aux \'etoiles observ\'ees
par Light et al., nous ne pr\'edisons pas les \event s courts.  
C'est naturel!  La masse totale de ces \'etoiles dans le bulbe n'est
que 10 milliard de masse solaire.  Mais, on 
croit que la masse totale dynamique est plut\^ot
20 milliard.  Il y a donc s\^urement d'autres \'etoiles dans le bulbe,
plus faibles que celles observ\'ees par Light et al.

	15) La solution la plus simple est d'\'etendre la fonction mesur\'ee
par Light et al.\ vers les petites masses 
avec celle du disque mesur\'ee par Hubble.  
Malheureusement,
\c ca ne marche pas non plus.  C'est aussi normale car
avec cette fonction de masse le bulbe n'aurait que
14 milliard de masse solaire, pas 20 milliard.

	16) On peut supposer que la fonction de masse du bulbe n'est pas
le m\^eme que celle du disque.  On peut choisir plut\^ot une loi de
puissance de Salpeter, comme Zhao et al., mais seulement pour les
masses moins qu'une demi-masse solaire,  ce qui est acceptable parce qu'il n'y
a aucune d'observations de ces \'etoiles.  Cette solution est meilleure
mais elle n'est pas parfaite.

	17)  Une autre solution est d'utiliser la fonction de masse de Light
et al.\ et \'etendue par Hubble pour les \'etoiles faibles, et d'ajouter 
6 milliard de masse
solaire de naines brunes d'une masse de 0,08 masse solaire.  
La masse totale est alors de
20 milliard, en accord avec estimations dynamiques, 
et la distribution observ\'ee
de $t_e$ est presque reproduite.

	18) EROS et MACHO ont tous deux observ\'e des candidats
\event s de microlentille vers le LMC.  EROS a men\'e deux
exp\'eriences, l'une en prenant quelques douzaines de poses chaque nuit dans le
m\^eme petit champ, et l'autre en faisent une longue pose chaque nuit
dans un grand champ.
La premi\`ere exp\'erience recherche des objets sombres de petite masse
entre $10^{-7}$ et $10^{-4}$ masse solaire.  Pour de tels objets,
on attend beaucoup d'\event s, mais chacun est tr\`es court.  On compare 
l'amplification du pic le plus grand avec le deuxi\`eme pic.  Si on simule
des \event s de microlentille par Monte Carlo, le rapport du deuxi\`eme pic
au premier pic est presque toujours proche de 0.  Cependant, pour les
donn\'ees il est presque toujours proche de 1, parce que ces objets sont
presque tous des \'etoiles variables qui se r\'ep\'etent.  Un certaine nombre
d'\event s subissent cette \'epreuve avec succ\`es, mais pour la plupart,
ils ont des amplifications tr\`es petites et sont donc
probablement des \'etoiles
variables rares.  A la fin, il n'a rest\'e que 5 candidats, et ce sont tous
des \'etoiles tr\`es brillantes qui constituent seulement 1\%
de toutes les \'etoiles, et elles sont en g\'en\'eral variables.
Il n'y a donc aucun candidats r\'eels.

	19) L'autre exp\'erience EROS recherche des objets sombres de masse
entre $10^{-4}$ et une masse solaire.  
Pour ces objets, on attend moins 
d'\event s, mais chacun dure de quelques jours \a
quelques semaines.  EROS a trouv\'e deux candidats montr\'es ici.
Apr\`es avoir publi\'e ces r\'esultats, EROS a d\'ecouvert que l'\'etoile 
source de l'un d'entre eux est une \'etoile binaire \a \'eclipse.  Ici, on voit
l'\event\ de microlentille superpos\'e \a la courbe de lumi\`ere de la
binaire \a \'eclipse.  Les \event s de ce type sont tr\`es int\'eressants et
je les discuterai dans la quatri\`eme le\c con.  On peut creindre
que l'\event\ ne soit pas une microlentille, mais il soit plut\^ot un transfer
de masse entre les deux \'etoiles.  Quant \a moi, je penche pour une
microlentille.

	20) Cependant, les poses ont \'et\'e prise sur des plaques 
photographique,
et la photom\'etrie \'etait donc mauvaise.  Pour cette raison,
EROS affirme seulement avoir
d\'etect\'e des `candidats', pas des `\event s'.  EROS n'a mis
qu'une limite sup\'erieure \a la quantit\'e d'objets sombres.  
EROS commencera une nouvelle
exp\'erience en mai o\`u juin qui utilisera une grande cam\'era de CCD.
L'exp\'erience de plaques
a eu de bonne efficacit\'e pour $t_e$ de quelques jours \a presqu'une ann\'ee.
Pour un halo compos\'e seulement de Machos de masse entre $10^{-7}$ 
et une masse solaire, on s'attend \a d\'etecter plus de 3 \event s
entre les deux exp\'eriences d'EROS.  On peut donc \'eliminer la plupart
des mod\`eles avec ces donn\'ees.  Mais la limite sup\'erieure
(en bleu) pr\`es d'une demi-masse solaire est tr\`es mauvaise \a cause des
deux candidats.  La limite sup\'erieure donn\'ee  par MACHO en analysant
ses donn\'ees de la premi\`ere ann\'ee (en rouge) est aussi mauvaise ici, 
et pour la m\^eme raison: il y avait des candidats.  Apr\`es la deuxi\'eme
ann\'ee, MACHO a dit que ses candidats ont \'et\'e vraiment des microlentilles
et que la densit\'e du halo a \'et\'e donc mesur\'ee, ici en vert.

	21) L'exp\'erience de MACHO utilise un cam\'era de CCD de 0,5
degr\'e carr\'e.  L'efficacit\'e est similaire \a celle d'EROS.  

	22) La selection de l'exp\'erience MACHO utilise
deux crit\`eres principaux: l'amplification maximum doit \^etre plus
grande que 1,75 et le $\Delta\chi$-deux doit d\'epasser 500.  C'est \a dire
que la courbe de 
microlentille doit \^etre beaucoup meilleur qu'une ligne
droite.  Il y a aussi d'autres crit\`eres de s\'election.  Par exemple, 
on a pu d\'eterminer que l'un de ces
cercles ouverts est une supernova.  Les deux autres peuvent \^etre r\'eels,
mais les \'etoiles source ont disparu apr\`es la fin des \event s.
Ces \event s n'ont \'et\'e remarqu\'es que parce qu'ils \'etaient en cours le
jour o\`u on a cr\'e\'e la liste des \'etoiles source.
Quatre de ces \event s (en rouge) ont de grandes amplifications et un
(en vert) est une lentille binaire.  

	23) Cette carte du LMC montre les positions des \event s, et aussi
les champs de MACHO.  Si la plupart des 
\event s \'etait dus \a des lentilles
dans le LMC, ils seraient fortement concentr\'es dans la
surface o\`u la densit\'e des \'etoiles est la plus grande.  En fait,
les \event s sont distribu\'es comme les \'etoiles.  Pour cette raison,
on pense que les lentilles sont dans le halo de la Voie Lact\'ee.  
N\'eanmoins, la lentille binaire est presque certainement dans le LMC. 

	24)  Ici, je vous montre les courbes de lumi\`ere pour les \event s.
Le premier est appel\'ee ``l'\event\ plaqu\'e or''.  Il est tr\`es beau,
n'est-ce pas?

	25) Le num\'ero 5 a une amplification de 40.

	26) MACHO pense que le num\'ero 10 n'est peut-\^etre pas
un \event\ de
microlentille, mais il a \'et\'e gard\'e parce qu'il a subi tous les \'epreuves
avec succ\`es.  Le num\'ero 9 semble un chaos affreux.  Mais c'est \a cause
d'un d\'efaut dans le CCD.  Quand on \'elimine toutes les mauvaises poses, 
la courbe de lumi\`ere devient celle d'une lentille binaire parfaite.

	27) Remarquons qu'il y a deux observations au passage de la premi\`ere 
caustique.
On peut donc calculer le temps de passage de la source \a travers
la caustique.  Puisqu'on conna\^\i t le rayon angulaire de la source
(par la loi de Stefan), on peut d\'eterminer la vitesse angulaire de la
lentille.  Cette vitesse angulaire correspond bien \a celle 
d'une lentille dans le LMC, mais elle est 10 fois plus petite que la valuer
moyenne des
lentilles dans le halo de la Voie Lact\'ee.  On peut aussi voir un
autre maximum, ici. 

	28) Dans la premi\`ere le\c con, je vous ai montr\'e que les maxima de
ce type se produisent quand la source passe pr\`es d'un 
point de rebroussement.

	29) Les \event s de MACHO impliquent qu'une fraction $f =0,5$ 
d'un halo standard est sous forme d'objets sombres, et que la masse
typique est proche d'une demi-masse solaire.  Mais, les erreurs sont grandes.
Par ailleurs, on ne conna\^\i t pas la forme du halo, et si on choisait une 
autre forme, les r\'esultats changeraient.

	30) On sait qu'au moins l'une des lentilles est dans le LMC, mais 
combien parmi les autres y sont aussi?  Certaines des lentilles, sont elles
dans le disque de la Voie Lact\'ee?  De toutes les mod\`eles du halo,
lequel est correcte?  Pour analyser proprement les \event s de microlentille, 
il faut obtenir plus d'informations.  Une m\'ethode, que nous avons d\'ej\a
vue appliqu\'ee aux \event s du bulbe, est d'analyser la distribution
des dur\'ees, $t_e$.  C'est la liste des \event s observ\'es jusqu'\a 
maintenant par MACHO (en rouge) et EROS (en bleu).  Les $t_e$ sont ici.
Cet \event\ ci est la source binaire et celui-l\a est la lentille binaire.
Tous les $t_e$ sont calcul\'es en supposant que la source n'est 
m\'elang\'ee avec aucune autre \'etoile.  En fait, il y a un indication 
de m\'elange
pour 4 des \event s et pour deux d'entre eux, l'effet de ce m\'elange
sur l'estimation de $t_e$
est grand.  La distribution de $t_e$ est consistente avec la distribution
obtenue si toutes les lentilles ont la m\^eme masse.  Mais les masses peuvent
aussi \^etre diff\'erentes.  Par ailleurs, les $t_e$ ne nous dissent rien sur
les positions des lentilles; sont elles dans le disque, le halo, ou le LMC?

	31) Une autre m\'ethode consiste \a anlayser la distribution des 
\event s dans le ciel.  Il y a dix minutes, j'ai utilis\'e cette m\'ethode 
pour d\'eduire
que les lentilles ne sont pas toutes dans le LMC.  Ici, je montre la 
distribution des \event s de MACHO observ\'es vers le bulbe.  On pr\'edit des 
distributions diff\'erentes pour les mod\`eles diff\'erents.

	32) Cependant, Cheongho Han et moi-m\^eme avons montr\'e
qu'il faut observer presque mille \event s pour pouvoir distinguer entre les
mod\`eles du bulbe.

	33) Pour comprendre les \event s observ\'es vers le LMC, la
question la plus importante est: o\`u sont les lentilles.  La m\'ethode
la plus efficace pour r\'epondre \`a cette question est de mesurer les 
vitesses projet\'ees sur le plan de l'observateur ou les
vitesses angulaires (aussi appel\'ees `movements propres') de toutes les
lentilles observ\'ees. La vitesse projet\'ee est simplement le rayon d'Einstein
projet\'e divis\'e par $t_e$.  Rappellons que le rayon projet\'e est mesur\'e
en utilisant la parallaxe.  
De la m\^eme fa\c con, la vitesse angulaire est le rayon angulaire
d'Einstein, $\theta_e$, divis\'e par $t_e$.
Cette table montre que ces quantit\'es sont
tr\`es diff\'erentes pour le disque, le halo, et le LMC.  Par exemple,
les vitesses projet\'ees des lentilles du halo sont typiquement
300 km/s, et celles du LMC sont quelques milliers de km/s.  Les
vitesses angulaires du halo sont 10 \a 50 fois plus grandes que celles
du LMC.  En fait, j'ai utilis\'e une mesure de vitesse angulaire
pour d\'eduire que la lentille binaire est dans le LMC.  Malheureusement,
il est beaucoup plus facile de mesurer les vitesses angulaires des
lentilles dans le LMC que dans le halo.  On n'apprend donc
presque rien sur la distribution des lentilles en mesurant les vitesses
angulaires.

	34) Je vous ai dit pendant la premi\`ere le\c con qu'on peut
mesurer les rayons d'Einstein projet\'es en utilisant un satellite, et 
par m\^eme occasion conna\^\i tre la vitesse projet\'e.

	35) Cependant, la situation est un peu plus compliqu\'ee que je ne vous
l'ai dit.  Je vous ai montr\'e ces deux courbes de lumi\`ere, 
et je vous ai dit qu'on peut reconstruire la g\'eom\'etrie de l'anneau
d'Einstein comme ci.  En fait, puisqu'on conna\^\i t seulement les valeurs
absolues des param\`etres d'impact $\beta$, on ne sait pas si la source
est vue du m\^eme c\^ot\'e de la lentille par la terre et par le satellite, 
o\`u qu'elle est vue de l'autre c\^ot\'e.  
Il y a donc 4 g\'eom\'etries possibles pour
l'anneau d'Einstein.  Deux parmi ces quatre ne pr\'esentent qu'un petit
probl\`eme parce qu'elles influent seulement sur la direction de la lentille.
Pour les deux autres, il y a aussi une d\'eg\'en\'erescence de $\Delta x$ qui 
affecte directement l'estimation du rayon d'Einstein et donc 
celle de la vitesse projet\'ee.

	36) N\'eanmoins, on peut utiliser la diff\'erence entre les vitesses
de la terre et du satellite pour r\'esoudre cette d\'eg\'en\'erescence.
La vitesse relative entre le satellite et la terre est a peu pr\`es
30 km/s et elle est dans la direction du soleil. A cause de cette diff\'erence,
la dur\'ee de l'\event\ $t_e$ n'est pas exactement la m\^eme quand il est
vu du satellite ou de la terre.  

	35) Si la source \'etait vue de la terre et du
satellite du m\^eme c\^ot\'e de la lentille,
le rayon d'Einstein projet\'e serait grand, la vitesse projet\'ee serait aussi
grande, et la faible vitesse relative entre le satellite et la terre
n'aurait pas beaucoup d'importance.  La terre et le satellite mesureraient
presque le m\^eme $t_e$.  Par contre, 
si la source \'etait vue de l'autre c\^ot\'e, le rayon d'Einstein
projet\'e et la vitesse projet\'ee seraient petits, la vitesse relative serait
importante, et les $t_e$ mesur\'es par le satellite et la terre
serait diff\'erents.

	37) Thomas Boutreux et moi-m\^eme avons d\'etermin\'e la
fraction de tous les \event s pour lesquels on peut casser cette 
d\'eg\'en\'erescence.
Si la distance entre la terre et le satellite \'etait au moins 0,5 unit\'es
astronomiques, et la masse de la lentille est plus grande que $10^{-2}$ masse 
solaire,
la d\'eg\'en\'erescence serait cass\'e pour la plupart des \event s 
vers le LMC.

	36) Le LMC est proche du p\^ole de l'\'ecliptique.  Pour cette raison,
la g\'eom\'etrie est relativement simple.  Elle est plus compliqu\'ee pour
le bulbe qui est seulement \a quelques degr\'es de l'\'ecliptique.  Pendant
une partie de l'ann\'ee, la s\'eparation projet\'ee entre le satellite
et la terre est tr\`es petite m\^eme quand la s\'eparation physique est
grande.  Pendant l'autre partie de l'ann\'ee, c'est la vitesse relative
qui est petite.  

	38) La fraction des \event s pour lesquels on peut casser la 
d\'eg\'en\'erescence, d\'epend 
donc fortement de la p\'eriode de l'ann\'ee, de la position
relative par rapport \a l'\'ecliptique,

	39) de la masse, et

	40) de la s\'eparation entre le satellite et la terre.  N\'eanmoins,
Scott Gaudi et moi-m\^eme avons montr\'e qu'on casserait
la d\'eg\'en\'erescence pour la plupart des \event s vers le bulbe si les 
sources \'etaient g\'eantes.

	41) Cependant, pour les \event s en la direction du bulbe, 
un satellite par lui-m\^eme ne nous donnerait pas assez d'informations.
On sait bien que beaucoup d'\event s dans cette direction sont dus \a des
\'etoiles.  Quelle est l'origine des autres?  C'est la question la plus 
int\'eressante.  J'ai d\'ej\a dit que beaucoup de ces \event s
sont probablement dus \a des naines brunes.  Pour confirmer 
cette hypoth\`ese,
et pour \'etudier ces objets, il faut d\'eterminer leur masses et leur
distances.  Comme je vous l'ai dit pendant la premi\`ere le\c con, ce serait 
possible
si on mesure le rayon d'Einstein projet\'e en utilisant un satellite et
le rayon angulaire d'Einstein, $\theta_e$.  On peut mesurer les 
petits $\theta_e$ en utilisant les effets de taille fini de source, et les 
grands $\theta_e$
par interf\'erometrie.  Ici, je montre la fraction des \event s
dus \a des sources g\'eantes, o\`u on peut mesurer $\theta_e$.  On devrait
pouvoir mesurer une trentaine d'\event s par an.

	42)  Les \'etoiles connues n'expliquent ni les \event s observ\'es
vers le bulbe ni ceux en direction du LMC.
Pour le bulbe, la profondeur optique observ\'ee n'est pas beaucoup
plus grande que
celle pr\'edite en tenant compte des \'etoiles connues.  Le probl\`eme 
principal est que les \'etoiles connues ne pouvent pas produire les
\event s courts qui sont observ\'es.  Peut-\^etre y a-t-il beaucoup de
naines brunes dans le bulbe.  Pour le LMC, $\tau_{\rm obs}$ est beaucoup
plus grandes que $\tau_*$.  La meilleure estimation des masses de ces
objet est une demi-masse solaire.  Les objets connus de cette masse
sont des naines blanches et des naines rouges.  Cependant, si le halo
\'etait compos\'e de ces objets, on pourrait les voir en utilisant Hubble.
Les observations nous mettent en face de grands myst\`eres.  Pour
les r\'esoudre, il faudrait lancer un satellite et faire en m\^eme temps
des nouvelles observations de la terre.

\chapter{La m\'ethode des pixels}

	1) Le sujet des deux premi\`eres le\c cons a \'et\'e la microlentille
classique.  Toutes les fois quand je parlais d'un ``\event '', je voulais
dire une amplification d'une \'etoile que l'on peut vraiment voir.  Tous
les \event s observ\'es en direction du bulbe ou du LMC sont de ce type.
En fait, je vous ai dit que MACHO a \'elimin\'e deux \event s parce
que les \'etoiles source avaient disparu apr\`es la fin des \event s.
Si ces deux \event s sont r\'eels, ils repr\'esentent des
amplifications d'\'etoiles non r\'esolues.  MACHO (ou EROS) ne
recherche pas des \event s de ce type, et donc il n'a pas pu les garder
parmi ses candidats.  Cependant, le bulbe de la Voie Lact\'ee et les 
Nuages
de Magellan sont les seules galaxies qui contiennent beaucoup d'\'etoiles
r\'esolues.  Si on veut rechercher des \event s de microlentille en
direction d'autres galaxies, on doit cr\'eer une m\'ethode pour
d\'etecter des amplifications des \'etoiles non r\'esolues.  Cette m\'ethode
est ``la m\'ethode des pixels''.

	2)  Si les \'etoiles d'une galaxie ne sont pas r\'esolues,
en premi\`ere approximation la galaxie appara\^\i t uniforme.  C'est \a dire
que le flux dans chaque pixel (ou dans chaque arcsec) est le m\^eme.  Ce
chiffre, le flux dans un angle solide, est appel\'e la brillance
de surface, $\Sigma$.  La brillance de surface ne d\'epend pas de la
distance de la galaxie.  
Si on la mettait \a une distance deux fois plus grande, il y aurait
quatre fois plus d'\'etoiles dans chaque pixel, mais chaque \'etoile
serait quatre fois moins brillante.  Le flux re\c cu par un pixel 
ne changerait donc pas.  
Ici, on voit que $\Sigma$, qui est d\'efini en termes du flux et
de l'angle solide, peut \^etre \'ecrite en termes de la densit\'e 
num\'erique et de 
la luminosit\'e des \'etoiles.  Puisque ces nombres ne d\'ependent
pas de la distance,  $\Sigma$ n'en d\'epend pas non plus.  Pour simplifier, 
j'ai suppos\'e que toutes les \'etoiles ont la m\^eme luminosit\'e, $L_*$,
et donc la m\^eme flux $F_*=L_*/4\pi d^2$.
Cependant, ce r\'esultat est toujours valable.

	3) M\^eme si la brillance de surface est invariable,
les {\it fluctuations} de la brillance de surface (FBS) changent selon
la distance.  Ici,
je montre la m\^eme galaxie vue \a trois distances diff\'erentes.  Dans la
figure sup\'erieure, il y a, en moyenne, 16 \'etoiles par pixel.  Cependant
\a cause des fluctuations de Poisson, certains pixels sont plus peupl\'es,
certains autres moins.  Ici, il y a $16\pm 4$ \'etoiles
par pixel.  On ne voit pas les \'etoiles, mais seulement le flux total
des \'etoiles dans chaque pixel.  Ces flux int\'egr\'es varient de
4 par rapport \a 16, soit 25\%.  
Si la galaxie \'etait 2 fois plus proche, il n'y aurait
que 4 \'etoiles dans chaque pixel, mais chaque \'etoile serait 4 fois
plus brillante.  La brillance de surface serait la m\^eme, comme nous
l'avons vu, mais les fluctuations serait plus fortes.  Ici, il y a
$4\pm 2$ \'etoiles et les fluctuations sont donc 50\%.  Si la galaxie \'etait
beaucoup plus proche, $n_*$, la densit\'e des \'etoiles par pixel, deviendrait
beaucoup plus petite que un.  Le nombre des \'etoiles serait donc 0 ou 1.
Autrement dit, les \'etoiles seraient r\'esolues!  En r\'ealit\'e,
ce n'est pas la taille du pixel qui est importante, mais celle du
disk de seeing, $\Omega_\psf$.  Je d\'efinirai pr\'ecisement $\Omega_\psf$
dans quelques instants, mais c'est essentiellement l'angle solide couvert
par l'image d'une source ponctuelle.
L'abr\'eviation ``psf'' vient de point spread function, qui est anglais pour
la fonction d'extension d'une source ponctuelle.  Un disque de seeing,
re\c coit en moyenne un flux de $\Sigma\Omega_\psf$, d\^u \a
$n_*\pm$ la racine carr\'ee de $n_*$ \'etoiles.  Le flux de chaque \'etoile
est $F_*$.  La fluctuation,
$\Delta\Sigma/\Sigma$ est donc la racine carr\'ee de $F_*/\Sigma\Omega_\psf$.

	4) Ici, je vous montre les fluctuations de la brillance de surface
de la grande galaxie dans Androm\`ede, M31.  Dans chaque disque de seeing,
il y a approxitivement $16\pm 4$ \'etoiles brillantes.  Les disques de
seeing contenant 20 \'etoiles apparaissent blancs et ceux contenant 
12 \'etoiles apparaissent noirs.

	5) En fait, des \'etoiles n'ont pas toutes les m\^emes luminosit\'es.  
Le contenu en \'etoile d'une galaxie est plut\^ot d\'ecrit par une fonction 
de luminosit\'e,
$\phi$.  N\'eanmoins, on peut d\'efinir une luminosit\'e de fluctuation,
$L_*$, le rapport du deuxi\`eme moment de $\phi$ \a son premier moment.
Les fluctuations de la brillance de surface seraient exactement les m\^emes
si toutes les \'etoiles avaient une luminosit\'e $L_*$.  Pour une
fonction de luminosit\'e typique, $L_*$ est 100 
luminosit\'es solaires ce qui correspond \a une magnitude absolue = $-1$.
Ici, je donne la d\'efinition pr\'ecise de $\Omega_\psf$ en termes de la 
fonction d'extension
d'une source ponctuelle, $\Psi_\psf$.  Si la taille du disque de seeing est, 
par
exemple, $2''$, $\Omega_\psf$ est approxitivement $\pi\times 2^2$ 
arcsec carr\'e.
On peut utiliser la magnitude de fluctuation pour mesurer les distances
des galaxies.  D'abord, on mesure les fluctuations, la brillance de la
surface, et la taille de la psf.  Puis, on calcule le flux re\c cu d'une
\'etoile ayant la magnitude de fluctuation, $F_*$.  Si l'on conna\^\i t
la luminosit\'e d'une telle \'etoile, $L_*$,  on peut en d\'eterminer
la distance.  Les distances de quelques centaines de galaxies ont \'et\'e
mesur\'ees de cette mani\`ere.  La magnitude de fluctuation est aussi tr\`es
utile pour comprendre la m\'ethode des pixels.  Par exemple, on peut
calculer $n_*$, le nombre des \'etoiles ayant la magnitude de fluctuation
dans chaque disque de seeing.  En utilisant des param\`etres typique, on 
trouve que $n_*$ dans le bulbe galactique est beaucoup moins grand que un et 
celui du Grand Nuage de Magellan est un peu moins grand que un.  Par contre, 
$n_*$ dans M31 est
sup\'erieur \a un et pour M87, ce chiffre est tr\`es grand.  En d'autre terms,
le bulbe nous pr\'esente des champs bien r\'esolus, les \'etoiles de M31
sont pour la plupart non r\'esolues, et M87 n'est pas
r\'esolue du tout.  Le Grand Nuage de Magellan est int\'eressant parce que
ses \'etoiles sont r\'esolues, mais pas bien.  En r\'esum\'e, la magnitude de
fluctuation peut d'abord servir classifier des champs.

	6) Pourquoi veut-on rechercher les \event s de microlentille
en direction de M31?  Premi\`erement, on peut mesurer la densit\'e des
objets sombres d'une galaxie autre que la Voie Lact\'ee.  
Si elle \'etait confirm\'ee, la d\'ecouverte
d'objets sombres dans le halo de la Voie Lact\'ee serait r\'evolutionaire.
Il est donc tr\`es important de se demander si
ces objets existent partout ou non.  Deuxi\`emement, M31 est peut-\^etre une
galaxie meilleure que la Voie Lact\'ee.  On peut rechercher des \event s
de microlentille sur plusieurs 
lignes de vis\'ees, vers le bulbe de 
M31 et aussi vers son disque.
La densit\'e des objets sombres devrait \^etre diff\'erente
sur chaque ligne de vis\'ee.  On devrait ainsi pouvoir reconstruire la 
distribution du halo en analysant ces observations.  
Par contraste, il n'y a que trois lignes de vis\'ee
pour la Voie Lact\'ee, le bulbe et les deux Nuages de Magellan.  Par ailleurs,
est-on absoluement s\^ur que les \event s observ\'es par MACHO et EROS
dans le LMC sont r\'eels?  Comment sait-on que ce ne sont pas des \'etoiles
variables tr\`es rares.  Une m\'ethode pour distinguer sans ambigu\"\i t\'e 
entre
les \event s de microlentille et les \'etoiles variables est de lancer un
satellite.  Les \event s de microlentille appara\^ \i traient 
diff\'erents si on les voyait
d'un satellite, mais les \'etoiles variables para\^\i traient semblables.  
Par contre, pour M31, il y a un test plus simple: la profondeur optique
vers la partie lointaine doit \^etre beaucoup plus grande que celle vers la
partie proche parce que M31 est inclin\'ee fortement
par rapport \a notre ligne de vis\'ee et l'on traverse beaucoup
plus du halo dans la premi\`ere direction que dans la deuxi\`eme.  
Cependant, M31 n'est pas meilleure
pour tout.  Ses \'etoiles sont beaucoup plus faibles que celles du
bulbe et du LMC.  Les observations sont plus difficiles et donc on tire
moins d'informations de chaque \event.  Une autre raison possible de
rechercher des \event s vers M31 est 
d'avoir une quatri\`eme ligne de vis\'ee \a travers du halo de la Voie 
Lact\'ee.  Cependant, comme
les \event s se produisent dix fois plus souvent
dans M31 elle-m\^eme que ceux dans la
Voie Lact\'ee, il n'est pas possible de d\'eterminer s\'epar\'ement
la profondeur optique
de la Voie Lact\'ee \a cette direction, sauf si l'on observe les \event s
avec un satellite.  Pour finir, on peut rechercher des objets sombres
plus legers que $10^{-6}$ masse solaire du halo de la Voie 
Lact\'ee.  Leurs rayons angulaires d'Einstein sont tellement petits 
qu'ils n'amplifient qu'une petite partie d'une source dans le LMC.  Puisque
M31 est 16 fois
plus loin que le LMC, les rayons angulaires de ses \'etoiles sont
16 fois plus petits, et l'amplification est efficace pour les lentilles
d'une masse qui est $16^2$ fois plus petites. 

	7)  Que mesure-t-on?  Je vous ai dit que pour une microlentille
classique, la courbe de lumi\`ere d\'epend de 3 param\`etres, $t_0$,
$\beta$, et $t_e$.  Bien s\^ur, il y a un quatri\`eme param\`etre, le flux
sans amplification, $F_0$.  Habituellement, on ne parle pas de $F_0$
comme un param\`etre parce qu'il est bien connu gr\^ace aux observations avant
l'\event.  Cependant, il n'y a pas seulement un quatri\`eme param\`etre mais
aussi un cinqui\`eme.  La lumi\`ere de la source n'est pas n\'ecessairement 
due \a une seule \'etoile.  La source peut-\^etre est une 
\'etoile binaire et dont le compagnon n'est pas amplifi\'e. 
La lentille elle-m\^eme peut aussi contribuer au flux.  
Dans tous les cas, un autre nombre intervient, 
le flux non amplifi\'e, $B'$.  Si $B'$ est connu, on peut \'ecrire
le flux comme \c ca, 
mais en g\'en\'eral, on ne conait que le flux total apr\`es 
l'\event, $B'+F_0$.  On doit donc \'ecrire le flux comme \c ca.  Puisque
$B$ ou $B'+F_0$ est bien connu, l'expression n'a effectivement que 
4 param\`etres, $F_0,\ t_0,\ \beta$ et $t_e$.  Pour la m\'ethode des
pixels, c'est la m\^eme chose.  On ne mesure que la diff\'erence entre
le flux amplifi\'e d'une \'etoile et le flux
avant l'\event.  La formule est donc exactement la m\^eme.  Cependant,
3 effets rendent difficiles les observations de microlentille sur des
\'etoiles non-r\'esolues.  D'abord, les \'etoiles sont faibles.  Par
exemple, le flux d'une \'etoile dans le LMC est 250 fois plus fort que celui de
la m\^eme \'etoile dans M31.  Deuxi\`emement, il y a un fond d\^u aux
autres \'etoiles non r\'esolues.  En r\'esum\'e, pour les champs d'\'etoiles 
non r\'esolues, le signal est plus petit et le bruit est plus grand.
Troisi\`emement, la comparison des images entre elles pr\'esente des 
difficult\'es pratiques dont je parlerai bient\^ot.
Il n'est souvent possible observer que d'\event s qui sont fortement 
amplifi\'es et
ces \event s ont n\'eanmoins un faible rapport signal sur bruit.  La
cons\'equence fr\'equente est une certaine d\'eg\'en\'erescence.
L'amplification est bien approxim\'ee par $1/x$, et le flux peut donc
\^etre \'ecrit comme ceci.  On voit 
que cette expression n'a que 3 param\`etres,
$t_0$ que nous avons d\'ej\`a vu, $F_{\rm max}$, le flux au maximum, et
$t_\eff$, la dur\'ee effective de l'\event.  On ne peut donc mesurer que
ces 3 param\`etres.  En particulier, on ne peut d\'eterminer ni
le flux $F_0$, ni la dur\'ee $t_e$ ind\'ependamment.

	8)  Comment fait-on la soustraction?  On a une image de r\'ef\'erence
avec les fluctuations de la brillance de surface (ou FBS)
montr\'ees en rouge.  L'image courante contient 
les m\^eme FBS, mais aussi 
le flux suppl\'ementaire (en bleu) d'une \'etoile amplifi\'ee.  
Pour mesurer le flux, on aimerait soustraire une image de l'autre
et obtenir une PSF parfaite!  Cependant, on rencontre quelques
probl\`emes.  D'abord, l'image n'est pas la fonction de variables
continues que j'ai trac\'e.  Elle contient 
des pixels discrets.  Heureusement, il y a des \'etoiles
r\'esolues de la Voie Lact\'ee dans l'image.  On peut les utiliser pour
aligner les deux images et pr\'edire le flux de l'image de
r\'ef\'erence qui
serait tomb\'e dans les pixels de l'image courante, si elle avait \'et\'e
d\'eplac\'ee une fraction de pixel.  Si les pixels sont beaucoup plus
petits que la PSF, \c ca marche.  Un autre probl\`eme est d'aligner 
la photom\'etrie des deux images.  Il y a en fait 2 probl\`emes.  
Premi\`erement, le fond de ciel change \a cause de, par exemple, la lune.
Deuxi\`emement, l'extinction atmosph\'erique change \a cause des
conditions m\'et\'eorilogiques et de la position de la galaxie dans le ciel.
Ce sont des probl\`emes classiques de l'astronomie que l'on r\'esoud 
normalement par une transformation lin\'eaire entre les images
bas\'ee sur les flux mesur\'es des \'etoiles.  Cependant, 
la solution en ce cas est plus difficile parce qu'il n'y a pas beaucoup 
d'\'etoiles
r\'esolues.  Jusqu'\a maintenant, on a quelques bonnes solutions mais aucune 
n'est parfaite.  Le probl\`eme le plus difficile est d'aligner les
PSFs.  Chaque image est une convolution de la PSF avec les sources ponctuelles
des \'etoiles dans la galaxie.  Si les PSFs de deux images
sont diff\'erentes, l'images
sont aussi diff\'erentes.  Il faut ramener les deux images au m\^eme
syst\`eme de PSF.  La encore quelques bonne solutions existent mais 
il faut continuer \a travailler \a ce probl\`eme.

	9) Le bruit introduit par la soustraction doit \^etre de m\^eme ordre 
ou plus petit que le bruit de photons.  
La magnitude de fluctuation est utile pour comprendre cette limite.  
Je d\'efinis le ``bruit-$\gamma$'' comme la racine carr\'ee du nombre de 
photons qui tombent dans une PSF au cours d'une pose.  
Ici, $\Sigma\Omega_\psf$ est
le flux dans la PSF, $t_{\rm pos}$ est la dur\'ee de la pose,
et $\alpha$ est le taux de d\'etection de photons par unit\'e de flux
arrivant sur le t\'elescope.
Le chiffre $N_{\gamma,*}$ est le nombre de photons d\'etect\'e au cours
d'une pose, venant d'une \'etoile ayant 
la magnitude de fluctuation.  Par exemple,
pour une pose de M31 avec un t\'elescope de 1 m\`etre et de dur\'ee 20 minutes,
ce chiffre est de l'ordre de mille.  Le rapport du bruit d'alignement
g\'eom\`etrique au bruit de photons est donn\'e par la racine carr\'ee de
$N_{\gamma,*}$ fois
ce rapport ci.  Par exemple, si la taille de la PSF, $\sigma_\psf$ \'etait 
\a peu pr\`es un pixel et si l'erreur de l'alignement, 
$\delta\theta$ 
\'etait quelques pourcents d'un pixel, le bruit g\'eom\`etrique serait 
suffisamment petit pour que le rapport soit moins grand que 1.  
De la m\^eme fa\c con, si l'erreur sur l'extinction \'etait de quelques 
pourcent, le bruit photom\'etrique serait aussi suffisamment petit.  
Si on ne faisait rien pour ramener les deux images au m\^eme syst\`eme
de PSF, le rapport de bruit de PSF au bruit statistique serait donn\'e
par cette formule, o\`u $\Delta\sigma$ sur $\sigma$ est la diff\'erence
relative entre les tailles des deux PSFs.  Si on essayait ramener les
deux images au m\^eme syst\`eme, par exemple, en convoluant une des images,
et s'il restait encore une diff\'erence $\Delta\sigma$-PSF entre les 
deux images, le bruit de PSF serait donn\'e par la m\^eme formule.
Pour \'eliminer effectivement
le bruit de PSF, il faut ramener les PSFs des deux images au m\^eme syst\`eme
\a mieux quelques pourcents aussi.

	10) Ces images, prises par Tomaney et Crotts, illustrent bien 
le probl\`eme de l'alignement de la PSF.
Ici, une image de M31 montre des fluctuations caract\'eristiques.
Au-dessous, on en a soustrait l'image de r\'ef\'erence apr\`es avoir bien
ramen\'e les deux images au m\^eme syst\`eme de PSF.  La plupart partie
de l'image diff\'erence est presque compl\`etement uniforme, 
comme un champ sans \'etoiles.
Il y a un probl\`eme ici \a cause d'une \'etoile satur\'ee, mais ce n'est pas
important.  A droite, on montre autre partie de la m\^eme image ainsi que
l'image apr\`es soustraction.  Malheureusement, la PSF 
de l'image originale dans
cette partie n'est pas la m\^eme que celle de la partie de gauche.
En cons\'equence, la soustraction est mauvaise.  Les fluctuations de la
brillance de surface sont 
r\'eduites, mais elles ne sont pas \'elimin\'ees. Si une \'etoile
avait vari\'e dans cette partie, on ne la remarquerait gu\`ere.  En fait,
une \'etoile a vari\'e au centre du petit carr\'e.  Quand on ramene
cette section de l'image au m\^eme syst\`eme de PSF que l'image de 
r\'ef\'erence, l'\'etoile devient vraiment visible.

	11) Il y a deux types d'\event s qu'on peut d\'etecter en utilisant
la m\'ethode des pixels.  Si les erreurs sont beaucoup plus petites que le flux
de la source (sans amplification), on peut voir l'\event\ m\^eme quand $A-1$ 
est beaucoup moins grand que 1, quand la source a quitt\'e l'anneau 
d'Einstein.  Dans ce cas, on peut d\'eterminer le temps que la source passe
\a l'int\'erieure de l'anneau d'Einstein.  C'est \a dire que l'on conna\^ \i t
$t_e$.  Les \event s de ce type sont donc appel\'es \event s
``semi-classique''.  On obtient les m\^emes informations sur ces \event s 
qu'on obtient sur des \event s classiques sur des \'etoiles r\'esolues.  Si
les erreurs sont du m\^eme ordre que le flux de la source, on ne peut voir
l'\event\ que quand la source est profondament \a l'int\'erieur de 
l'anneau d'Einstein.  On ne peut pas mesurer les $t_e$ de ces \event s
\a cause de la d\'eg\'en\'erescence dont je vous ai parl\'e.  Puisque
leurs amplifications sont n\'ecessairement tr\`es grandes, je les appelle
``les \event s pic''.  Dans la direction de M31, on peut voir des \event s
semi-classique sur des sources brillantes et des \event s pic sur des sources
faibles.  Par contre, vers M87, on ne peut voir que
des \event s pic.

	12) Ici, je montre la profondeur optique et le taux des
\event s qu'on attend \a la direction du centre de M31.  Les \event s
pr\'edits
sont presque tous dus \a des lentilles situ\'ees dans le bulbe de M31.
La profondeur optique au centre est \a peu pr\`es $10^{-5}$.  Jusqu'\a 
maintenant, toutes les observations ont \'et\'e prises dans ce champ.
Puisque ce champ a la profondeur optique la plus grande, c'est
le meilleur pour tester la m\'ethode.
Par contre, pour d\'etecter le halo proprement dit, il faut observer dans une
direction distant de quelques kpc du centre de M31, o\`u on attend peu
d'\event s dus \a des lentilles du bulbe de M31.

	13) Ceci est la distribution des \event s en fonction du rapport
signal sur bruit.  Notons qu'il y a 5 fois plus d'\event s avec un
rapport plus grand que 20 que d'\event s avec un rapport plus grand que 80.
Pour mesurer $t_e$, il faut que le rapport signal sur bruit soit au moins
80.  Notons aussi que le taux d'\event s de la Voie Lact\'ee est 10
fois plus petit que celui de M31.  Il faudrait donc pouvoir distinguer entre 
les \event s de M31 et de la Voie Lact\'ee, si on voulait d\'eterminer la 
profondeur optique du halo
de la Voie Lact\'ee dans cette direction.  On aurait pu esp\'erer qu'il serait
possible d'utiliser 
les dur\'ees $t_e$ pour les distinguer, mais malheureusement,
les deux distributions sont presque les m\^emes.
 
	14) Cette courbe de lumi\`ere est observ\'ee vers M31 par AGAPE.
Elle est compatible avec un \event\ de microlentille, mais elle est
probablement une \'etoile variable de longue p\'eriode.  Pour distinguer
entre les deux, il faut observer le champs beaucoup plus longtemps que
la dur\'ee de l'\event.  Cette courbe de lumi\`ere illustre aussi deux
id\'ees importantes.  D'abord la figure sup\'erieure montre toutes les
observations sans distinction de seeing.  Dans l'autre figure, 
les points de mauvais seeing sont \'elimin\'es.  Si on voulait utiliser les
donn\'ees de mauvais seeing, il faudrait corriger ces mesures.  Deuxi\`emement,
les graphes du bas montrent la surface de M31 au voisinage du pixel
central.  Au maximum, l'\'etoile est r\'esolue ou demi-r\'esolue.  Par contre,
la bosse dans l'image du minimum est peut-\^etre l'\'etoile mais
elle peut aussi \^etre une fluctuation de brillance due \a d'autres
\'etoiles non r\'esolues.

	15) On peut choisir entre deux attitudes envers les \'etoiles 
variables.
Elles constituent un fond g\^enant ou un tr\'esor scientifique.  De toute 
fa\c con, il faut les reconna\^\i tre pour s\'eparer les \event s de 
microlentille
r\'eels.  MACHO a trouv\'e plus d'\'etoiles variables dans le LMC que
tous les autres observateurs du monde.  L'analyse de ces variables est
tr\`es int\'eressante, mais je n'ai pas le temps de la discuter.  Ici,
je montre deux variables des donn\'ees AGAPE.  Une des \'etoile est
probablement une cephe\"\i d et l'autre est une nova.  On peut utiliser la
m\'ethode des pixels pour rechercher des variables non resolues dans
les amas stellaires de la Voie Lact\'ee aussi bien que dans les galaxies
ext\'erieures.

	16) R\'efl\'echissons \a deux \event s pic hypoth\'etiques.
Le flux de l'\'etoile de source du premier est la moiti\'e de celui de l'autre,
mais son param\`etre de l'impact est quatre fois plus petit.
On peut calculer que le flux maximum du premier \event\ est deux fois
plus grand,  et que sa dur\'ee effective est quatre fois plus petite.
Le rapport signal sur bruit pour une mesure proche du pic est deux 
fois plus grand pour le premier \event\ que pour le second, 
mais il y a quatre fois moins d'observations.  En effet,
les rapports signal sur bruit totaux des deux \event s ont \'et\'e choisis 
identiques.
Supposons que le premier \event\ soit tout juste perceptible.  Alors, tous les
\event s avec la m\^eme source et un $\beta$ plus petit serait aussi
d\'etectable.  Puisque le deuxi\`eme \event\ a le m\^eme rapport signal
sur bruit que le premier, il est aussi tout juste perceptible.  Si une
\'etoile de source est deux fois plus brillante qu'une autre, on peut
donc d\'etecter des \event s avec un $\beta$ quatre fois plus 
grand.  C'est \a dire qu'on peut voir quatre fois plus des \event s.
En r\'esum\'e, le nombre d'\event s est proportionel au flux au carr\'e.

	17) Le taux de d\'etection d'\event s sur toutes les \'etoiles
de source est donc une int\'egrale de la fonction de luminosit\'e $\phi$
multipli\'e par $\beta_{\rm max}(F)$, la valeur de $\beta$ qui rend un \event\
tout juste perceptible.  Puisque $\beta_{\rm max}$ varie comme
$F^2$, cette int\'egrale est la m\^eme que celle qu'on utilise
pour calculer la magnitude de fluctuation.  Le taux d'\event s $\Gamma$
est donc donn\'e par cette formule en termes de $\Gamma_{\gamma,*}$, le taux 
de d\'etection de photons venant d'une \'etoile ayant la magnitude de 
fluctuation.  Les
autres nombres dans cette formule sont $Q_{\rm min}$, le rapport
signal sur bruit  minimum 
exig\'e pour d\'etecter un \event, $\tau$ la profondeur
optique, et $N_{\rm res}$, le nombre d'\'el\'ements de r\'esolution dans
l'image de 
CCD.  C'est la troisi\`eme utilisation de la magnitude de fluctuation.
La premi\`ere a \'et\'e pour distinguer les galaxies qui ont des \'etoiles
r\'esolues de celles qui n'en ont pas.  La deuxi\`eme a \'et\'e pour
comprendre le rapport du bruit syst\'ematique au bruit statistique.
Revenons \a la troisi\`eme.  Notons
en particulier que le taux est proportionel \a la profondeur
optique mais ne d\'epend pas des dur\'ees des \event s.  Pour
les microlentilles classiques, on d\'etermine $\tau$ en mesurant les $t_e$,
alors que pour les \event s pic on peut d\'eterminer $\tau$ m\^eme si les $t_e$
sont inconnus.  C'est un r\'esultat tr\`es important parce qu'il montre
que la m\'ethode des pixels nous donne des informations quantitatives sur
la densit\'e des objets sombres ce qui n'est pas \'evident.

	18) Il est beaucoup plus difficile d'observer des \event s
dans M87 que ceux de M31 parce que M87 est 20 fois plus loin et
ses \'etoiles sont donc 400 fois plus faibles.  N\'eanmoins il serait
tr\`es int\'eressant les observer.  Pourquoi?  Les exp\'eriences MACHO
et EROS trouvent une masse totale de 100 milliard de masses solaires pour les
objets sombres du halo de la Voie Lact\'ee.  Ce chiffre est du
m\^eme ordre que la masse visible du disque et du bulbe de la Voie Lact\'ee.
Alors, je mod\'elise la formation de notre galaxie ainsi.  Un milliard
ann\'ees apr\`es le Big Bang, une moiti\'e du gaz (en bleu) est 
transform\'ee en Machos (en rouge).  
Quelques milliards d'ann\'ees plus tard, le gaz s'effondre et
il devient un proto-disque et un proto-bulbe.  Puis, le disque et le
bulbe produisent des \'etoiles.  Dans le mod\`ele standard de la formation d'un
amas de galaxies, un grand nombre des galaxies commencent \a se former dans
un r\'egion d'espace de grande densit\'e.
Alors, imaginons qu'une galaxie se formant pr\`es d'un amas de galaxies soit
similaire \a la Voie Lact\'ee.  
Une moiti\'e du gaz de cette galaxie est aussi transform\'ee en
Machos (en rouge).  Cependant, avant que le reste du gaz puisse s'effondrer,
la galaxie traverse le gaz chaud de l'amas et perd tout son gaz.
On s'attend donc \a ce que les masses sous forme de gaz et sous forme de 
Machos sont \a peu pr\`es les m\^emes dans un amas.  Puisqu'\a peu pr\`es
20\% de la masse d'une amas est sous forme de gaz, peut-\^etre 20\% est sous
forme de Machos.

	19)  Aujourd'hui, on peut utiliser la m\'ethode des pixels pour
observer M87 au centre de l'amas de Virgo.  S'il y a des Machos dans le halo
de l'amas, on peut les voir.  Cependant une \'etoile de la magnitude de 
fluctuation dans Virgo est extr\^ement faible.   Pour l'observer 
malgr\'e le bruit, elle doit \^etre bien amplifi\'ee.  S'il y avait
beaucoup de bruit \a cause de, par exemple, un ciel brillant ou un
grand disque du seeing, on pourrait en principe attendre des \event s
tr\`es rares avec des param\`etres d'impact tr\`es petits.  
Malheureusement,
\a cause de la taille finie des sources, il y a une amplification maximum,
m\^eme si le param\`etre d'impact est tr\`es petit.  
Alors, on doit chercher le meilleur
seeing possible ainsi que le fond de ciel le plus faible.  
Pour faire ces observations on doit donc utiliser Hubble.  
Si la fraction de Machos dans le halo de 
Virgo est plus que 20\%, on devrait voir plus de 3 \event s par jour. 

	20)  On peut aussi appliquer la m\'ethode des pixels \a des champs
o\`u beaucoup d'\'etoiles sont r\'esolues tel que le bulbe de la Voie Lact\'ee.
Cela peut semble tr\`es stupide puisqu'il y a beaucoup d'\'etoiles
bien r\'esolues dans ce champ.  Il y a aussi beaucoup
d'\'etoiles non r\'esolues, mais on doit faire un grand effort pour les
observer.  Pourquoi s'en soucier?  Envisageons les \event s de
microlentille extr\^emes avec des amplifications $A=$ 200 ou plus.
Il y a deux exigences. D'abord, le param\`etre d'impact doit \^etre
plus petits que 1/200.  Deuxi\`emement, le rapport du rayon de la source
au rayon angulaire d'Einstein doit aussi \^etre plus petit que 1/200.
Ces \event s sont donc tr\`es rares et les sources doivent \^etre
tr\`es petites.
Cependant, ces \event s extr\^emes seraient tr\`es pr\'ecieux si on pouvait
les observer.

	21)  Pourquoi?  D'abord, on pourrait d\'eterminer les parallaxes de
ces \event s en mesurant la courbe de lumi\`ere depuis 2 positions sur
terre.  Rappelons qu'un \event\ serait situ\'e diff\'eremment dans 
l'anneau d'Einstein
si on l'observait d'une autre position.  Le d\'eplacement $\Delta x$ en 
unit\'e du rayon d'Einstein est
\'egal \a la distance entre les observateurs divis\'ee par le rayon d'Einstein
projet\'e.  Puisque le rayon projet\'e est de quelques unit\'es astronomiques,
l'id\'ee est naturelle que l'autre observateur soit un satellite \a une 
distance proche d'une UA. L'effet sur la courbe de lumi\`ere est \a peu pr\`es 
$\Delta x$ qui serait donc de
l'ordre unit\'e.  Si les observateurs \'etait tous deux sur
terre, $\Delta x$ serait \a peu pr\`es $10^{-5}$ et la taille de l'effet
serait du m\^eme ordre.  Seuls les th\'eoriciens incorrigibles peuvent
penser qu'un effet aussi petit est discernable.  C'est une id\'ee de Holz et 
Wald, et effectivement Wald est l'auteur fameux d'un livre tr\`es
abstrait sur la relativit\'e g\'en\'erale.  En fait, la taille de
l'effet n'est pas $\Delta x$, mais $\Delta x/\beta$.  Pour les \event s
extr\^emes, ce chiffre est de l'ordre 1\%.  Peut-on mesurer des effets
de 1\%?

	22) \c Ca d\'epend.  Il est difficile de mesurer 1\% d'un petit
flux, mais c'est possible si le flux est grand.
Pour les \event s vus dans la direction du bulbe, l'exigence que
$\theta_e$ soit 200 fois plus grand que $\theta_*$ implique que
la source soit une \'etoile d'un rayon solaire.  Les \'etoiles de ce type
ont des magnitudes corrig\'ees de rougissement $I_0= 19$ et
des magnitudes apparentes $I=20$ \a 22.  Cependant, quand elles sont
amplifi\'ees, elles atteignent des magnitudes corrig\'ees $I_0=13$.
Elles sont donc 10 fois plus brillantes que les sources g\'eantes.  Il n'est
pas tr\`es difficile de mesurer pr\'ecisement le flux d'\'etoiles si 
brillantes.
Malheureusement, il est tr\`es difficile de les trouver.  Avant l'\event,
il y a une de ces \'etoile dans chaque carr\'e de 3 arcsec de c\^ot\'e.
Autrement dit, ces \'etoiles sont non r\'esolues et elles n'entrent pas
dans le catalogue des \'etoiles qu'on suit pour rechercher des \event s
de microlentille classique.  N\'eanmoins, un jour avant le maximum, l'\'etoile
serait reconnaissable facilement en utilisant la m\'ethode des pixels
parce qu'elle serait une nouvelle \'etoile de la brillance
d'une g\'eante.  Apr\`es, on peut
l'observer en utilisant les m\'ethodes classiques.  Comme je vous l'ai dit,
on peut d\'eterminer la parallaxe de ces \event s depuis la terre.  
Par ailleurs,
puisque la lentille passe pr\`es de la source, on peut aussi mesurer
le rayon angulaire d'Einstein en utilisant la photom\'etrie optique/infrarouge.
On d\'eterminerait donc la masse, la distance, et la vitesse de la lentille
pour peut-\^etre 75 \event s par an.  Pourquoi, a-t-on besoin d'\'etoiles
non r\'esolues pour mesurer la parallaxe depuis la terre?  D'abord, les
\event s extr\^eme sont tr\`es rares, et il y a beaucoup plus d'\'etoiles
non r\'esolues que r\'esolues.  Mais il y a un autre probl\`eme plus
fondamental.  Les \'etoiles r\'esolues sont plus brillantes et donc, plus
grandes.  Si le param\`etre d'impact \'etait tr\`es petit, la lentille
passerait \a l'int\'erieur de la source et la courbe de lumi\`ere
serait aplatie.  Les deux observateurs sur terre verraient le m\^eme
maximum et ne pourraient donc pas d\'eterminer une parallaxe.

	23) En conclusion, la recherche des microlentilles n'est pas
limit\'ee aux \'etoiles r\'esolues.  En soustrayant une image de l'autre,
on peut regarder des \event s de microlentille m\^eme quand les \'etoiles
sont non r\'esolues.  Il est plus difficile de faire ces observations parce
que les fluctuations de la brillance de surface produisent un bruit
syst\'ematique.  Il est possible de reduire ce bruit au-dessous du bruit
de photons, et deux groupes (y compris AGAPE ici) ont bien avanc\'e en cette
direction.
Jusqu'\a maintenant, la m\'ethode des pixels n'ont \'et\'e appliqu\'ee 
qu'\a M31.  Cependant, on peut appliquer la m\^eme m\'ethode pour 
rechercher les objet sombres dans l'amas de Virgo.  Dans ce cas, on ne verrait
que les \event s pic, o\`u la source est tr\'es amplifi\'ee.  Paradoxallement,
on peut aussi l'appliquer au bulbe de la Voie Lact\'ee m\^eme quand il y a
beaucoup d'\event s sur des \'etoiles r\'esolus dans ce champ.  
Les \event s qu'on peut y 
d\'etecter en utisilant la m\'ethode des pixels sont tr\`es int\'eressants
parce qu'on peut d\'eterminer leur masse et leur distance.  Dans la
quatri\`eme lecon, je vous dirai comment on peut utiliser la m\^eme 
m\'ethode pour \'etudier l'histoire de la formation des \'etoiles.

\chapter{L'avenir des microlentilles}

	1) On a commenc\'e les exp\'eriences de microlentille pour
rechercher des objets sombres de la mati\`ere cach\'ee, et elles nous 
donnent vraiment de bonnes m\'ethodes pour \'etudier ces objets.  Cependant,
comme nous avons d\'ej\a vu, on peut aussi appliquer ces m\'ethodes
pour examiner d'autres questions.  Par exemple, dans la deuxi\`eme
le\c con, j'ai discut\'e des microlentilles observ\'ees vers le bulbe
de la Voie Lact\'ee.  Ces lentilles sont des \'etoiles faibles, des
\'etoiles brillantes, et aussi peut-\^etre des naines brunes.  On peut
combiner ces r\'esultats avec d'autres informations pour reconstruire
la fonction de masse de tous les objets du disque et du bulbe, lumineux
et sombres.  Cependant, cette application est tr\`es proche des
utilisations originales des microlentilles.  En fait, quand Paczy\'nski
et Griest ont sugg\'er\'e cette id\'ee il y a 5 ans, personne ne l'a
pens\'ee tr\`es r\'evolutionnaire.  Mais, presqu'imm\'ediatement, quelques
personnnes ont reconnu que si on peut d\'etecter des \'etoiles normales
en utilisant les effets de microlentilles, on peut aussi rechercher
d'\'eventuelle plan\`etes associ\'ees.  
Bien que le sujet de plan\`etes est tr\`es diff\'erent
de celui de la mati\`ere cach\'ee, leur d\'etection posent le m\^eme 
probl\`eme:  rechercher des objets tr\`es faibles.  N\'eanmoins,
il est possible d'appliquer les effets de microlentilles \a des probl\`emes
compl\`etement diff\'erents.  Par exemple, on peut les utiliser pour 
mesurer les vitesses transverses de galaxies ou les vitesses de rotation
d'\'etoiles.  L'effet de microlentille peut devenir une m\'ethode 
astronomique tr\`es g\'en\'erale comme la photom\'etrie de tavelures 
ou l'interf\'erometrie.
Aujourd'hui, je vais vous parler de l'avenir des microlentilles, des
applications qui peuvent sembler un peu bizarres, mais qui indiquent
des possibilit\'es de cette m\'ethode encore tr\`es jeune.

	2) Il est tr\`es difficile de rechercher des plan\`etes.  Pourquoi?
D'abord, elles sont tr\`es faibles.  M\^eme si la terre r\'efl\'echissait
toute la lumi\`ere du soleil, elle n'aurait qu'une luminosit\'e de $10^{-10}$
luminosit\'e solaire.  Le nombre est similaire pour jupiter parce qu'il
a une surface 100 fois plus grande, mais il re\c coit 25 fois moins du
flux solaire.  N\'eanmoins vous pouvez facilement ``d\'ecouvrir'' venus
et jupiter ce soir en utilisant les instruments classiques de 
Copernicus et Hipparchos: vos yeux.  
Une raison est que ces plan\`etes sont proches.
Si elles \'etaient \a une distance de 10 pc,  elle seraient de la trenti\`eme 
magnitude, tout juste perceptible par les t\'elescopes les plus grands.
Mais, le probl\`eme principal est que l'on peut regarder jupiter et venus
pendant la nuit quand la terre obstrue la lumi\`ere du soleil.  Les
plan\`etes des autres \'etoiles sont s\'epar\'ees de
moins de $1''$ de leur ``soleil'',
et on devrait donc \'eliminer cette lumi\`ere tr\`es forte si on voulait les
d\'etecter directement.
Jusqu'\a maintenant, la m\'ethode principale pour d\'etecter les plan\`etes
est de rechercher leurs effets gravitationnels sur leur \'etoile centrale.
Malheureusement, Newton nous a dit que pour chaque force, il y a une autre
force \'egale et oppos\'ee.  Puisque la vitesse de la terre n'est que
30 km/s et le rapport de la masse de la terre \a celle du soleil est
seulement 1/300.000, la vitesse du soleil due au mouvement de la terre
est 10 cm/s.  Il n'est pas possible de d\'etecter un effet
si faible.  De m\^eme fa\c con, le soleil ne se d\'eplace que
de 500 km \a cause du mouvement de la terre.  On pourrait facilement
d\'etecter un mouvement de cet ordre si la plan\`ete 
\'etait dans le syst\`eme solaire, mais \a la distance de 10 pc, 
le mouvement propre ne serait pas perceptible.  Les
chiffres sont beaucoup plus grands pour jupiter, et il y a quelques
exp\'eriences qui recherchent des plan\`etes de cette masse en mesurant
les changements
des vitesses radiales ou des positions angulaires des \'etoiles centrales.

	3) Les microlentilles nous donnent une m\'ethode profondemment 
diff\'erente pour rechercher des plan\`etes.  Les perturbations de l'effet de
microlentilles dues \a une plan\`ete ne sont pas petites.  Ils sont de l'ordre 
de l'unit\'e. 
La probabilit\'e absolue d'un \event\ plan\'etaire est tr\`es petite.
Par exemple, si chaque \'etoile avait un ``jupiter'', la profondeur
optique pour des ``jupiters'' vers le bulbe ne serait que $10^{-9}$.  Pour
les ``terres'' elle serait $10^{-11}$.  En effet, on ne trouvera jamais
des plan\`etes en utilisant les m\'ethodes de microlentilles normales.
Cependant, la probabilit\'e relative est beaucoup plus grande.  Si
on avait {\it d\'ej\`a} d\'etect\'e l'effet de microlentille produit par une
\'etoile, et si cette \'etoile avait une plan\`ete, la probabilit\'e
de d\'etecter la plan\`ete est na\"\i vement de 1\% \a quelques pourcents.
Par exemple, le rayon d'Einstein d'un ``jupiter'' est un trenti\`eme
de celui d'une \'etoile d'une masse solaire parce que le soleil est mille
fois plus massif que jupiter et le rayon d'Einstein varie selon
la racine carr\'ee de la masse.  La premi\`ere source ici produit
un \event\ de microlentille normal qui dure 30 jours.
Pendant cet \event, il y a un autre \event\ plan\'etaire d\^u au rayon
d'Einstein d'un ``jupiter'' (en vert).  Pour la plupart des \event s,
la source ne vient pas assez pr\`es du rayon d'Einstein de la plan\`ete et
on ne voit donc que l'\event\ normal.  Notons, m\^eme si l'\event\
plan\'etaire survient, on ne pourrait pas le voir sauf si on observait 
la source
tr\`es fr\'equemment.  L'\event\ de l'\'etoile dure quelques dizaines de
jours, mais l'\event\ plan\'etaire ne dure que 1 jour ou moins.  On
doit l'observer plusieurs fois par jour pour d\'etecter la plan\`ete.
En fait, le sch\'ema montr\'e ici est trop simple.  L'\'etoile et
la plan\`ete constituent ensemble un syst\`eme binaire.  Pour le comprendre
correctement, on ne peut pas analyser l'effet de la plan\`ete en utilisant
le petit rayon d'Einstein circulaire d'un ``jupiter'' isol\'e.  N\'eanmoins,
les r\'esultats d'une analyse propre sont qualitativement les m\^emes.

	4) Cette figure illustre avec r\'ealisme un \event\ plan\'etaire.
La dur\'ee de l'\event\ stellaire est 30 jours et pour une grande partie de sa
dur\'ee il semble bien \^etre un \event\ normal.  Cependant, environs 7 jours 
apr\`es le pic, un autre \event\ survient.  Il a lui aussi une grande 
amplification, mais
il est tr\`es court, \a peu pr\`es un jour.  La g\'eom\'etrie de l'anneau
d'Einstein est un peu plus compliqu\'ee que celle indiqu\'ee dans la 
figure pr\'ec\'edent.  La source passe \a travers l'anneau d'Einstein de la 
lentille
stellaire et l'image de la source est perturb\'ee quand la source entre dans
cette zone ci.  Cependant, deux choses sont diff\'erentes de l'id\'ee
na\"\i ve pr\'esent\'ee avant.  D'abord, la zone de perturbation n'est pas un 
cercle comme
un petit anneau d'Einstein.  Deuxi\`emement, la plan\`ete n'est pas au
centre de cette zone.  Comment peut-on comprendre ces diff\'erences?

	5) Au commencement, imaginons une source et une lentille stellaire
mais sans plan\`ete.  Il y a deux images, $I_+$ et $I_-$ parce que
la lumi\`ere de la source est d\'eflechie par la lentille.  Puis, mettons
une plan\`ete exactement sur le chemin de la lumi\`ere de l'une des
images.  Naturellement, l'image est tr\`es perturb\'ee et l'amplification
est modifi\'ee aussi.  Si la plan\`ete reste l\`a, mais la source est
d\'eplac\'ee, il y a encore un effet, indiqu\'e ici en vert.  L'effet
reste grand si la source est d\'eplac\'ee une grande distance le long de l'axe
de l'\'etoile et de la plan\`ete, mais il diminue rapidement si la source
est d\'eplac\'ee \a l'autre direction.  Si la plan\`ete est sur le chemin
de l'autre image (celle \a l'int\'erieur de l'anneau d'Einstein), il y a 
aussi un effet, mais il est plus petit et la g\'eom\'etrie est un peu 
diff\'erente.

	6) Cette figure illustre le m\^eme principe mais avec plus 
d'exactitude.
Si la position de la plan\`ete est au dehors de l'anneau d'Einstein
\a $x_p=1,3$, la zone perturb\'ee est tr\`es \'eloign\'ee.  Le diamant
ici est la caustique.  Une lentille binaire produit toujours une caustique.
Les courbes plus \'epaises repr\'esentent des augmentations de l'amplification
de 5\%, 10\% etc, relatives au cas d'une lentille ponctuelle.  Les courbes
en traits fins repr\'esentent des diminutions.  Si la plan\`ete s'\'ecarte de
l'anneau d'Einstein, par exemple \a $x_p=2,2$,
les courbes deviennent plus circulaires, comme l'amplification due \a une
plan\`ete isol\'ee.  Si la plan\`ete est \a l'int\'erieur de l'anneau
d'Einstein, elle produit deux petites zones de perturbation.  

	7) Un \event\
est produit quand une source passe \a travers la structure due \a 
l'amplification.  La plupart des \event s plan\'etaires comme A et D semblent
\^etre des \event s normaux mais avec des dur\'ees tr\`es courtes.  Cependant,
il y a aussi des \event s ayant des caustiques comme B and C.  Remarquons que 
l'\event\ C a un autre pic parce que la source passe pr\`es d'un
point de rebroussement.  La g\'eom\'etrie de l'amplification est
la m\^eme que celle de la figure pr\'ec\'edent et l'\event\ D est le
m\^eme que celui de la figure initiale.

	8) Jusqu'\a maintenant, je supposais que le rayon d'Einstein
de la plan\`ete est beaucoup plus grand que le rayon de la source.  En fait,
ce n'est pas une bonne supposition pour les ``terres''.  On peut
d\'efinir $Q$ comme le carr\'e du rapport du rayon source au rayon
d'Einstein plan\'etaire.  Pour les ``jupiters'', $Q$ est plus petit
que 1 et le traitement pr\'ec\'edent marche bien.  La probabilit\'e d'un
\event\ plan\'etaire varie selon la racine carr\'ee de la masse plan\'etaire
sur la masse stellaire, et l'effet est de l'ordre 1.  Par contre, pour
une ``terre'', la probabilit\'e varie selon la taille de la source, et
pas selon la taille de l'anneau d'Einstein plan\'etaire.  Elle est donc
augment\'ee par la racine carr\'ee de $Q$.  Malheureusement, la grandeur de 
l'effet tombe d'un facteur $Q$ parce que seulement 
un petite zone de la source est ampfli\'ee.

	9)  Ici je montre une g\'eante du bulbe amplifi\'ee par une ``terre''
qui est \a la moiti\'e de la distance vers le bulbe.  Pour simplifier, j'ai
suppos\'e que la terre est isol\'ee.  C'est \a dire qu'elle est bien au dehors
de l'anneau d'Einstein de l'\'etoile.  La courbe \'epaise indique le rayon
de l'\'etoile non amplifi\'ee.  On voit que l'amplification est \a peu
pr\'es 5\%.  Cependant, il y a un autre effet.  Le bord (montr\'e en rouge)
est amplifi\'e plus que le centre.  Comme nous avons d\'ej\a vu dans la
premi\`ere le\c con, le bord d'une source g\'eante est plus rouge que
son centre.  

	10) On d\'etecterait donc ainsi un changement de couleur pendant
l'\event\ si on le mesurait en $V$ et $H$ en utilisant 
une cam\'era optique/infrarouge.
Le changement de couleur n'est que de quelques pourcents, mais il est
tr\`es important parce qu'il d\'emontre que la lentille r\'esourd la source
et donc que la lentille doit \^etre tr\`es petite: une plan\`ete.

	11) Pour d\'etecter des plan\`etes, on doit observer des \event s
plus courts que 1 jour, et on doit donc essayer de 
faire des observations fr\'equentes, 24 heures
par jour.  Bien s\^ur, c'est impossible en utilisant un seul observatoire.
Dans la premi\`ere le\c con, je vous ai montr\'e cette figure d'un
\event\ qui illustre les effets de taille finie de source.  
Notons qu'il y a approximativement
6 observations chaque jour.  Cet \event\ \'etait observ\'ee par GMAN
en utilisant 6 t\'elescopes autour du monde.

	12)  GMAN et un autre groupe PLANET recherchent des plan\`etes
en suivant attentivement des \event s vers le bulbe signal\'es par MACHO
et OGLE.  Ces donn\'ees viennent de 2 des 4 t\'elescopes de PLANET.
L'\event\ est une lentille binaire d\'ecouverte vers le bulbe par
MACHO.

	13)  La meilleure m\'ethode pour rechercher des plan\`etes est
de suivre un grand nombre de sources g\'eantes d\'ej\a amplifi\'ees
par des lentilles.  Les sources g\'eantes sont meilleures pour deux raisons.
Puisqu'elles sont brillantes, on les observe dans un temp plus court.  
En cons\'equence, on peut suivre plus d'\'etoiles en utilisant le m\^eme
t\'elescope.  Par ailleurs, elles ont des grands rayons angulaires, et la 
probabilit\'e de d\'etecter une ``terre'' est donc plus grande.  Pour
rechercher un grand nombre de g\'eantes, il faut observer une grande
partie du bulbe.  Jusqu'\a maintenant, les exp\'eriences n'observent que les 
champs avec les extinctions les plus faibles.  Mais, EROS 
observera peut-\^etre la plupart du bulbe
l'ann\'ee prochaine.  Si on observait le bulbe entier, on 
d\'ecouvrirait une centaine d'\event s de sources g\'eantes
par an.  Si chaque lentille avait
une plan\`ete s\'epar\'ee approximativement de un ou deux rayons d'Einstein,
et si l'on suivait ces \event s 
fr\'equemment en utilisant des t\'elescopes autour du monde, on
d\'etecterait 17 ``jupiters'' ou 3 ``terres'' par an.  On peut bien 
d\'eterminer le rapport de la masse plan\'etaire \a la masse stellaire qui est 
simplement le carr\'e du rapport de la dur\'ee de l'\event\ plan\'etaire
\a celle de l'\'etoile.  On peut aussi bien d\'eterminer la position 
projet\'ee de la plan\`ete en unit\'es du rayon d'Einstein.  C'est la valeur 
de $x$ \a l'instant du pic de l'\event\ plan\'etaire.  Puisqu'on peut toujours
estimer \a un facteur deux la masse de l'\'etoile et la taille de son rayon 
d'Einstein, on peut estimer avec la m\^eme pr\'ecision la masse plan\'etaire 
et la
s\'eparation physique projet\'ee entre l'\'etoile et la plan\`ete.  Pour
les \event s de ``terre'', la caustique passe presque toujours \a travers 
la source.  Comme nous avons d\'ej\a vu, on peut donc d\'eterminer
le rayon angulaire d'Einstein.  Si les \event s \'etaient observ\'es aussi
par un satellite, on mesurerait leur parallaxe.  Rappelons que le rayon
angulaire d'Einstein et la parallaxe ensemble nous donneraient la masse,
la distance, et la vitesse de la lentille.  On conna\^\i trait donc
la masse et la s\'eparation projet\'ee de la plan\`ete.

	14) Il y a beaucoup de m\'ethodes pour mesurer le rayon
angulaire d'Einstein.  On peut les diviser en 2 types.  Pour les
grands rayons, la meilleure m\'ethode est l'interf\'erometrie.  
Malheureusement, le premier instrument ne sera pa termin\'e avant l'an 2000.
Avant cela, on peut mesurer des grands rayons en utilisant des
occultations lunaires des \event s de microlentille.  La lune occulterait
l'une image quelques millisecondes avant l'autre et ...

	15) on pourrait voir un \'echantillon caract\'eristique comme
on voit pendant une occultation d'une \'etoile binaire.
La courbe verte indique le flux attendu si la source ne comportait qu'une 
image.  Puisque la source consiste en deux images, la courbe de lumi\`ere
(en noir) d\'evie.  On peut d\'eterminer la s\'eparation entre les images
en mesurant cette diff\'erence.

	16) Pour mesurer des petits rayons angulaires d'Einstein, il faut
utiliser la taille finie de la source.  Le probl\`eme principal est que
la source est ordinairement trop petite.  Plusieurs personnes ont essay\'e
d'esquiver ce probl\`eme en utilisant des solutions ing\'enieuses.  Une de
ces id\'ees nous conduira \a une application de microlentille 
compl\`etement nouvelle.  Il y a 5 ans, Griest \& Hu ont montr\'e que certains 
\event s ayant des sources binaires apparaissent tr\`es diff\'erents
des \event s normaux.  R\'ecemment, Cheongho Han a r\'ealis\'e que l'on peut
mesurer le rayon angulaire d'Einstein en utilisant cet effet.  Si la source est
consitu\'ee de deux \'etoiles similaires en masse et en luminosit\'e,
la binaire revient \a sa position originale apr\`es la moiti\'e de sa 
p\'eriode.  La p\'eriode des oscillations est donc la moiti\'e de celle de la
binaire.  Par contre, si une des \'etoiles est beaucoup plus brillante que
l'autre, les deux p\'eriodes sont les m\^emes.  Le cas moyen est aussi
possible.  On peut utiliser la courbe de lumi\`ere pour d\'eterminer 
le rapport du rayon d'Einstein \a la s\'eparation 
des \'etoiles binaires.  On peut alors d\'eterminer la s\'eparation
physique en faisant 
les observations de spectroscopie suivantes.  On profite de la
grande taille d'une source binaire.  Puisque la s\'eparation
entre les deux \'etoiles est beaucoup plus grande que le rayon d'une
source, il est beaucoup plus facile de reconna\^\i tre les effets de taille 
finie de source.  
Si la p\'eriode binaire est beaucoup plus longue que la dur\'ee 
de l'\event, il y a une d\'eg\'en\'erescence, la m\^eme que pour la
parallaxe.  On ne conna\^\i t pas si les deux sources sont vues du m\^eme
c\^ot\'e de la lentille ou non.  Cependant, si la dur\'ee de l'\event\ 
est au moins un dixi\`eme de la p\'eriode binaire, on peut souvent lever
cette d\'eg\'en\'erescence parce que les courbes de lumi\`ere pour les 
deux cas sont tr\`es diff\'erentes.

	17) Rappelons que la source d'un \event\ d'EROS est une
binaire \a \'eclipse.

	18) Une autre id\'ee pour \'etendre la taille finie de la source est 
due \a Dani Maoz.  Les raies spectroscopiques d'une \'etoile s'\'elargissent 
avec sa vitesse de rotation.  Un c\^ot\'e de l'\'etoile se rapproche vers
nous et
ses raies sont d\'ecal\'ees vers le bleu.  Les raies de l'autre c\^ot\'e
sont d\'ecal\'ees vers le rouge.  Habituellement, le r\'esultat est un
\'elargissement sym\'etrique.  Cependant, si le c\^ot\'e rouge est plus
proche de la lentille, les raies d\'ecal\'ees vers le rouge sont plus 
amplifi\'ees, et les centres des raies sont d\'ecal\'es syst\'ematiquement 
vers le rouge.  
Si on mesure le d\'ecalage vers le rouge et la vitesse de rotation
de l'\'etoile, on peut d\'eterminer le rayon d'Einstein angulaire.  Cet effet
est proportionnel \a l'inverse de la s\'eparation entre la lentille et la 
source.  Par
contre, l'effet photom\'etrique varie selon le carr\'e de la s\'eparation.
Si la vitesse de rotation est grande, on peut donc mesurer l'effet 
spectroscopique m\^eme si la s\'eparation est importante.

	19) Pour les \'etoiles du type A dans le LMC, les vitesses de 
rotation projet\'ees sont de l'ordre de 100 km/s et cette m\'ethode est tr\`es 
bonne.  Ici, je montre l'effet comme une fonction du temps pour diff\'erentes
g\'eom\'etries.

	20) Cependant, il n'est pas possible d'appliquer cette m\'ethode
aux g\'eantes du bulbe pour deux raisons.  D'abord, les vitesses de rotation
des g\'eantes sont tr\`es petites.  L'effet est donc aussi petit.  Plus
fondamentalement, pour utiliser cette m\'ethode, il faut mesurer la
vitesse de rotation de la source.  Ces vitesses seraient tr\`es int\'eressantes
mais personne ne les a jamais mesur\'ees.  On a mesur\'e les vitesses
de rotation des anc\^etres des g\'eantes, les \'etoiles de la s\'equence
principale.  On les a aussi mesur\'ees pour les descendantes des g\'eantes
sur la branche horizontale.  Mais il est tr\`es difficile de mesurer ce
chiffre pour les g\'eantes parce que leur vitesse de rotation est beaucoup
plus petite que les mouvements turbulents \a leur surface.  N\'eanmoins, on
aimerait beaucoup conna\^\i tre ces vitesses.  Puisque les vitesses de
rotation des anc\^etres et des descendantes sont toutes deux mesur\'ees,
on pourrait apprendre de l'\'evolution du mouvement angulaire des \'etoiles
si cette quantit\'e \'etait aussi mesur\'ee pour les g\'eantes.  
En particulier, on pourrait
distinguer entre deux mod\`eles des g\'eantes:  celui de la rotation du corps 
solide et celui o\`u le mouvement angulaire par unit\'e de masse est constant.
On peut ainsi r\'esoudre plusieurs probl\`emes d'\'etoiles incluant l'abondance
de lithium qui est importante pour la cosmologie.

	21)  Comment mesure-t-on la vitesse de rotation de g\'eantes?
Bien s\^ur, en utilisant des microlentilles.  Si la source passe pr\`es de la
lentille, le d\'ecalage vers le rouge des raies est donn\'e par cette
formule ci, o\`u $v$ sin$\,i$ est la vitesse de rotation projet\'ee
que l'on veut mesurer, et $\alpha$ est l'angle entre l'axe de rotation et
la ligne de la source \a la lentille.  La fonction $G(z)$ est montr\'ee
ici, o\`u $z$ est le rapport de la s\'eparation au rayon de la source.
Si la lentille passe a l'int\'erieur de la source ($z$ moins grand que 1),
on peut d\'eterminer $z$ en utilisant la variation du flux par rapport au flux
d'une source ponctuelle (en bleu) dont j'ai d\'ej\a discut\'e dans la 
deuxi\`eme le\c con.
On conna\^\i t donc $G(z)$.  On peut ainsi d\'eterminer sin $\alpha$
et $v$ sin $i$ s\'epar\'ement en faisant plusieurs observations pendant
l'\event.

	22) Je vais discuter d'une autre application \a un probl\`eme
compl\`etement diff\'erent: l'histoire des \'etoiles.  Pour faire
cette application il faut rechercher des \event s microlentille d'un million 
de quasars.  On s'attend \a d\'etecter une vingtaine d'\event s par an
dus \a des \'etoiles dans les galaxies.  Si $\Omega$-Machos, la densit\'e
de Machos relative \a la densit\'e critique de l'univers, \'etait
1\%, on aurait d\'etect\'e de l'ordre de 200 \event s par an dus \a ces objets.
Les Machos peuvent \^etre proches des galaxies comme les objets d\'etect\'es
par MACHO et EROS, ou ils peuvent \^etre dans l'espace intergalactique.

	23) Supposons que des \'etoiles (ou Machos) aient \'et\'e 
form\'ees aussit\^ot 
apr\`es le Big Bang, ce qui correspond \a $\alpha =0$ dans ces figures
o\`u $z$-quasar est le d\'ecalage vers le rouge de la source quasar.
La profondeur optique vers un quasar de $z=4$ serait beaucoup
plus grande que celle d'un quasar de $z=1$, parce que la ligne de vis\'ee
passe \a travers un plus grand nombre d'\'etoiles.  Si la formation des
\'etoiles \'etait uniforme dans le temps ($\alpha=1,5$), la profondeur optique
vers un quasar de $z=4$ ne serait pas beaucoup plus grande que celle vers un 
quasar de $z=1$ parce que peu d'\'etoiles auraient \'et\'e form\'ees \a grand
$z$.  On peut donc d\'eterminer le taux de formation
d'\'etoiles et de Machos en mesurant ce rapport.

	24) On peut aussi mesurer l'amplitude des mouvements particuliers
des galaxies, des d\'eviations du flot de Hubble.  Le soleil se d\'eplace
relativement au fond de $3^\circ$ K \a 300 km/s.  Si on observe un quasar 
perpendiculaire
\a notre direction de mouvement, les galaxies sur la ligne de vis\'ee 
paraissent se d\'eplacer vers l'autre direction.  Pour un quasar
parall\`ele \a notre direction de mouvement, il n'y a pas d'effet.  Si
les mouvements intrins\`eques des galaxies \'etaient petits, les mouvements
relatifs des galaxies vers les directions perpendiculaires (en rouge)
seraient beaucoup plus grands que ceux vers la direction parall\`ele
(en bleu).  Les \event s de
microlentille dus \a des \'etoiles vues dans des directions perpendiculaires
seraient plus courts et plus fr\'equents.  Par contre, si les mouvements
intrins\`eques \'etaient grands, le mouvement du soleil n'aurait pas beaucoup
d'importance, et les deux distributions appara\^\i traient \a peu pr\`es
les m\^emes.  Nous n'avons pas de bonne chance, car en ce moment, 
le mouvement du soleil autour la Voie Lact\'ee
est contraire \a celui de notre galaxie relativement au
fond de $3^\circ$ K.  Apr\`es cent millions d'ann\'ees ces mouvements seront
parall\`eles, et le mouvement du soleil relatif au fond de $3^\circ$ K sera
de 700 km/s, pas de 300.  L'exp\'erience sera donc beaucoup plus sensible \a
ce moment l\`a.

	25) Malheureusement, l'exp\'erience peut rencontrer quelques petits
probl\`emes.  D'abord, il faudrait regarder un million de quasars, mais on n'en
conna\^\i t que $10^4$.  Deuxi\`emement, on reconna\^\i t un \event\ de 
microlentille \a cause de la variation apparente de la source, mais les quasars
ont aussi des variations intrins\`eques.  Troisi\`emement, il faut
bien comprendre les observations tr\`es int\'eressantes de Rudy Schild.
Le premier probl\`eme n'est pas un probl\`eme du tout.  On peut regarder
un quart du ciel ($10^4$ degr\'es carr\'es) plusieurs fois par an
en utilisant un t\'elescope de 1 metre et une cam\'era de 2 degr\'es.
Il ne faut pas essayer de trouver des quasars. On ne recherche que
des variabilit\'es en utilisant la m\'ethode des pixels.  En fait, EROS
utilisera la m\^eme m\'ethode pour rechercher des supernovae.

	26) En principe, il n'est pas difficile de distinguer un
\event\ de microlentille d'une variation intrins\`eque d'un quasar.
Si le quasar variait, il illuminerait la r\'egion de raies larges,
et quelques mois plus tard, le gaz de cette r\'egion r\'e\'emettrait 
la lumi\`ere par fluorescence.  Une variation de flux des raies suivrait
la variation de flux du quasar.  Par contre, les flux des raies ne
varieraient pas \a la suite d'un \event\ de microlentille.  Malheureusement,
le monde ne contient pas assez de t\'elescopes pour faire des observations
spectroscopiques d'un million de quasars.  Cependant, on peut rechercher
une population de quasars avec une variabilit\'e faible et restreindre des 
observations spectroscopiques aux candidats de microlentille parmi
cette population.  On ne sait pas s'il existe une telle population, mais
des exp\'eriences \a venir nous donneront la r\'eponse \a cette question.

	27) Jusqu'\a maintenant, je ne parlais que de la th\'eorie.  En fait,
Rudy Schild avait observ\'e deux images du quasar 0957+561 depuis 15 ans.
Les deux images sont s\'epar\'ees par $6''$ et sont produites par une
MACROlentille, une galaxie interpos\'ee sur la ligne de vis\'ee.  Le quasar 
varie, et on peut donc mesurer le retard entre les deux images ce qui 
repr\'esente \a peu pr\`es 410 jours.  Si aucune des deux images n'etait 
affect\'ee par des microlentilles, leurs flux varieraient exactement de la 
m\^eme fa\c con.  La diff\'erence entre les images doit donc nous d\'evoiler 
des microlentilles.  On s'attend \a ce que la dur\'ee de l'\event\ $t_e$ soit 
30 ann\'ees pour une
microlentille d'une masse solaire.  En fait, on voit un \event\ de ce type.
Mais il y a aussi beaucoup d'autres structures dans les donn\'ees.  Par
exemple, ici je montre un \event\ de 30 jours observ\'e sur l'image A 
mais pas sur l'image B.  Si cet \event\ \'etait d\^u a une microlentille,
la masse serait \a peu pr\`es le carr\'e de 30 jours sur 30 ann\'ees, ou
$10^{-5}$ masse solaire.  Par ailleurs, elle n'amplifierait qu'une petit
zone du quasar puisque l'amplification est seulement de 4\%.  Rappelons
que les exp\'eriences MACHO et EROS ont fortement limit\'e la densit\'e
des objets de cette masse dans le halo de la Voie Lact\'ee.  S'il y avait
beaucoup d'objets de cette masse 
dans une autre galaxie, ce serait tr\`es int\'eressant.
Mais, ce n'est pas tout.  Les courbes de lumi\`ere des images A et B
sont en d\'esaccord pour beaucoup d'\'echelles de temps.  Par exemple, ici 
une courbe montre des variations d'une dur\'ee de un ou deux jours alors que
l'autre est relativement constante, et ici les courbes sont toutes deux
relativement constantes mais elles restent diff\'erentes de 2\%.
Ces effets ne sont pas grands, mais c'est tout de m\^eme difficile de 
comprendre pourquoi ils existent.  Avant de rechercher des \event s de 
microlentille parmi un million de quasars, il faut comprendre ces 
observations de quasar 0957+561 de Rudy Schild.

	28)  Les microlentilles donnent des espoirs aux projets 
d\'esesp\'er\'es.  Par exemple, elles rendent possible la mesure des vitesses
transverses de galaxies.  Habituellement, on ne pense jamais \a les
mesurer parce que c'est \'evidemment impossible sauf pour les galaxies
qui sont des satellites de la Voie Lact\'ee.  En g\'en\'eral, on d\'etermine
la vitesse transverse d'un objet en mesurant sa vitesse angulaire
(ou mouvement propre) et sa distance, puis en multipliant les deux.
La vitesse angulaire d'un satellite de la Voie Lact\'ee n'est qu'un
dixi\`eme d'arcsec par si\`ecle.  On a mesur\'e ce chiffre
pour le Grand Nuage de Magellan, mais c'\'etait tr\`es difficile.  La
vitesse angulaire de l'amas de Coma est mille fois plus petite.  Bien
s\^ur, on peut facilement d\'eterminer les vitesses radiales de galaxies
en mesurant leur d\'ecalage vers le rouge, mais c'est tr\`es difficile
de mesurer leurs vitesses transverses.

	29) Pourquoi veut-on mesurer les vitesses transverses de galaxies
(VTG)?  D'abord, au premier ordre la vitesse radiale ne nous dit rien sur
la vitesse particuli\`ere d'une galaxie, mais seulement sur sa distance.  
La vitesse radiale comprend deux composantes, la vitesse de flot de Hubble 
et la vitesse particuli\`ere, et la premi\`ere est en g\'en\'eral la plus 
grande.  Pour d\'eterminer la vitesse particuli\`ere,
on doit obtenir ind\'ependemment une estimation de la distance de la galaxie.
Puisque ces estimations ne sont pas tr\`es pr\'ecises, on n'obtient 
presque jamais
bonne mesure de la vitesse particuli\`ere.  En mesurant la vitesse
particuli\`ere des galaxies, 
on peut d\'eterminer la masse, par exemple, de l'amas de
Virgo.  Dans le mod\`ele simple expos\'e ici, les vitesses particuli\`eres 
(en vert) sont toutes dirig\'ees vers l'amas.   Nous ne mesurons que leur
composante radiale.  Les galaxies devant l'amas sont d\'ecal\'ees
vers le rouge et celle en arri\`ere sont d\'ecal\'ees
vers le bleu.  En r\'ealit\'e, les mesures sont mauvaises et on ne peut
faire qu'une estimation statistique de la masse.  En fait, pour faire
cette estimation,
il faut supposer une relation tr\`es simple entre les vitesses radiales et
les vitesses transverses.  On suppose que le flot de toutes les galaxies
n'a pas de rotation.  Beaucoup de nos id\'ees en cosmologie d\'ependent
de cette supposition, mais on ne pourrait 
l'\'eprouver qu'en mesurant les vitesses tranverses.

	30)  Deuxi\`emement, comme je vous l'ai dit, il faut mesurer la
distance d'une galaxie pour d\'eterminer sa vitesse particuli\`ere.  
L'erreur de cette mesure est proportionnelle \a la distance et devient
du m\^eme ordre que la vitesse particuli\`ere si la galaxie est \a une
distance de 100 Mpc.  On aimerait faire des mesures des vitesses qui
ne d\'ependent pas de la distance.  La solution: des microlentilles!

	31) D'abord, il faut trouver un \event\ de microlentille dans
une galaxie spirale.  Pour ces galaxies, on peut d\'eterminer les 
vitesses des \'etoiles relativement au centre de la galaxie en mesurant
sa courbe de rotation.  On conna\^\i t les distances de la galaxie
et du quasar en mesurant leur d\'ecalage vers le rouge.  Si la distance
de la galaxie \'etait 100 Mpc, le rayon d'Einstein projet\'e serait
quelques centaines de UA.  S'il y avait un satellite suffisamment loin
de la terre, on pourrait mesurer la parallaxe, et donc la vitesse
projet\'e de la lentille dans la galaxie.   On ne fera qu'une petite erreur
en corrigeant pour le mouvement de la lentille relativement \a la galaxie.
Deux probl\`emes interviennent.  D'abord, puisque le rayon d'Einstein
projet\'e est \a peu pr\`es 400 UA, l'effet de parallaxe ne serait pas
grand sauf si le satellite \'etait tr\`es loin.  Quand j'ai propos\'e cette
id\'ee, j'ai dit qu'il fallait mettre un t\'elescope de 1 m sur une orbite
comme Neptune.  L'effet est donc de l'ordre de 10\%.
Cependant, NASA a r\'ecemment commenc\'e \a discuter d'un t\'elescope de 8
m sur une orbite comme jupiter.  Dans ce cas, l'effet n'est que l'ordre de
1\%, mais on peut mesurer ce petit effet en utilisant un t\'elescope de
8 m.  On peut peut-\^etre mesurer 3 vitesses transverses par an.  Le
deuxi\`eme probl\`eme est le m\^eme que pour toutes les mesures de parallaxe:
la d\'eg\'en\'erescence.  Cependant, puisque l'\event\ dure 3 ans, 
on peut casser la d\'eg\'en\'erescence en utilisant le mouvement annulaire 
de la terre.

	32) Rappelons qu'une microlentille produit deux images ayant
les amplifications $A_\pm$ donn\'e par cette formule ci.  Habituellement,
on n'a pas besoin des deux formules mais seulement de leur somme.  Cependant,
si les deux image se font interf\'erence, il faut ajouter les racines 
carr\'ees des amplifications et non les chiffres eux-m\^emes.  Quand 
l'interf\'erence est contructive ou destructive, l'amplification
maximum ou minimum est donn\'ee par cette formule,
o\`u $x$ est la s\'eparation entre la source et la lentille en unit\'es
du rayon d'Einstein.  Le rapport du maximum au minimum est donc \'ecrit
comme \c ca.

	33) Les deux images n'arrivent pas \a l'observateur en m\^eme temps.
L'image plus proche de la masse arrive plus tard que l'autre \a cause
de deux effets.  D'abord, la lumi\`ere doit parcourir une distance plus
grande.  Deuxi\`emement, la vitesse de la lumi\`ere est plus lente dans
un potentiel gravitationnel.  Le retard entre les deux images est donn\'e
par cette formule ci, o\`u le param\`etre $\eta$ est un peu compliqu\'e.
Cependant, si $x$, la separation entre la source et la lentille en
unit\'es du rayon d'Einstein, est moins grand que 1, $\eta$ est approxitivement
\'egal \a $x$.  Le retard est donc simplement le temps n\'ecessaire pour
traverser un trou noir de la m\^eme masse, multipli\'e par quelques facteurs
de l'ordre 1.  Ici, $z_L$ est le d\'ecalage vers le rouge de la lentille.
Ce r\'esultat est valable pour les lentilles ponctuelles, mais les formules
pour les autres lentilles sont similaires.  Rappelons que le retard
de 0957 mesur\'e par Rudy Schild \'etait \a peu pr\`es d'une ann\'ee.
Cette mesure indique que la masse de la lentille est de l'ordre
de $10^{12}$ masses solaires en accord avec autres mesures de la masse de
cette lentille galactique.  Si la lentille n'avait qu'une masse
de $10^{-16}$ masse solaire, le retard serait 28 ordres de magnitude plus
petit, \a peu pr\`es $10^{-20}$ seconde.  Peut-on mesurer un aussi petit 
retard?  Oui!  Le retard produirait une diff\'erence des phases des deux image
$\Delta\phi$ donn\'ee par cette formule, o\`u $m_e$ est la masse de
l'\'electron.  Si la source \'emettait des rayons $\gamma$, ce retard
serait de l'ordre 1 et les deux images se feraient interf\'erence.

	34) En fait, la plupart des personnes croit que les sursauts de 
$\gamma$ sont des objets \a des distances cosmologiques.  Ces sources nous
permettent donc de rechercher des objets sombres d'une masse de $10^{-15}$ \a
$10^{-16}$ masse solaire.  Habituellement, le spectre d'un sursaut de
$\gamma$ appara\^\i t comme ceci, sans motif visible.  Mais, si une lentille
tr\`es petite \'etait dans la ligne de vis\'ee, on verrait un
motif d'interf\'erence.  L'interf\'erence est constructive
pour les fr\'equences o\`u le retard est un multiple entier de la
p\'eriode et destructive s'il est semi-entier.  Le rapport
du maximum au minimum nous donne la valeur de $x$, et l'intervalle 
d'energie entre les maxima nous donne le retard.  On peut donc
facilement calculer la masse, ou plut\^ot la masse fois ($1+z_L$).
Cette m\'ethode est utile pour rechercher des lentilles de 
$10^{-15}$ masse solaire et avec les rayons angulaires d'Einstein de
$10^{-15}$ arcsec.  Ces lentilles sont donc appel\'ees ``femtolentilles''.

	35) Une autre m\'ethode utilise aussi des sursauts $\gamma$
mais pour rechercher les objets de masse $10^{-7}$ \a $10^{-15}$
masse solaire.  Rappelons que les masses inf\'erieures \a $10^{-7}$ sont
trop l\'egeres pour amplifier des \'etoiles dans le LMC parce que leur
rayons angulaires d'Einstein sont plus petits que ceux des sources.
Cependant, pour une lentille et une source toutes deux d'une distance
cosmologique, le rayon d'Einstein projet\'e est une UA si la masse est
$10^{-7}$.  On peut donc faire l'exp\'erience suivante.  Observer des
sursauts $\gamma$ en utilisant deux t\'elescopes similaires mais 
s\'epar\'es par une UA.  S'il n'y avait pas de microlentille, les
flux re\c cus par les deux t\'elescopes devraient \^etre les m\^emes.
Par contre si une masse entre $10^{-15}$ et $10^{-7}$ masse solaire
se trouvait sur la ligne de vis\'ee d'un des t\'elecopes, le flux sur ce 
t\'elescope serait amplifi\'e, mais celui sur l'autre serait non
amplifi\'e parce que le rayon d'Einstein serait trop petit.  On peut
donc utiliser les effets de microlentille pour rechercher des objets de
toutes dont la masses se situe entre $10^{-16}$ et $10^6$ masses solaires.

	36)  En conclusion, l'avenir des microlentilles est 
brillante.  On peut les utiliser pour rechercher des plan\`etes.
Les ``jupiters'' produisent des effets d'ordre 1 et sont donc facilement
d\'etect\'es.  Les microlentilles nous donnent la seule m\'ethode pour
d\'etecter des ``terres'' en utilisant des observations sur notre terre.
Deux exp\'eriences ont d\'ej\a commenc\'e.  Les microlentilles ne
sont pas utiles seulement pour rechercher des objets sombres.  J'ai
discut\'e de trois autres applications, mesurer les vitesses de rotation
d'\'etoiles g\'eantes, mesurer les vitesses transverses de galaxies,
et \'etudier l'histoire de la formation des \'etoiles.  Pour finir,
les microlentilles, ou plut\^ot les femtolentilles, nous r\'ev\`eleront 
peut-\^etre des objets de la masse d'une com\`ete se trouvant de l'autre
c\^ot\'e de l'univers.

\singlespace
{\bf La reconnaissance}: Ces le\c cons sont \'ecrites et pr\'esent\'ees
pendant un s\'ejour d'un mois au Coll\`ege de France.  Je veux remercier
 Professeur Marcel Froissart et l'Assembl\'ee des Professeurs pour
m'inviter, et toutes les personnes du Coll\`ege pour faire mon s\'ejour
tr\`es agr\'eable.  En particulier, je veux remercie Jean Kaplan pour
corriger le fran\c cais de la plupart de ces le\c cons.  

\bye